\documentclass[reprint,superscriptaddress,amsmath,amssymb,aps,prx]{revtex4-2}
\usepackage{graphicx}
\usepackage{dcolumn}
\usepackage{bm}
\usepackage{hyperref}
\usepackage{xcolor}
\usepackage{physics}

\renewcommand\b[1]{{\bf  #1}}

\renewcommand\rm{\mathrm}

\renewcommand\dd{\mathrm{d}}

\begin{document}

\preprint{APS/123-QED}

\title{Thermalized buckling of extensible, semiflexible polymers}
\author{Richard Huang}
\author{David R. Nelson}
\affiliation{Department of Physics, Harvard University, Cambridge, Massachusetts 02138, USA}

\author{Suraj Shankar}
\affiliation{Department of Physics, Harvard University, Cambridge, Massachusetts 02138, USA}
\affiliation{Department of Physics, University of Michigan, Ann Arbor, Michigan 48109, USA}

\date{\today}

\begin{abstract}
The Euler buckling of rods is a long-studied mechanical instability, and it remains relevant to this day, as the constituent components in many biological and physical systems are linear polymers, such as microtubules or carbon nanotubes. At finite temperature, if a polymer is shorter than its persistence length, the polymer is semiflexible, and its elasticity remains rod-like. But polymers can also stretch due to their finite extensibility, which can couple to energetically cheap bending deformations in nonlinear ways when a load is applied to the system. We show how the interplay between thermal fluctuations and nonlinear elasticity dramatically modifies the Euler buckling instability for compressed semiflexible polymers in a fixed strain ensemble. We identify a Ginzburg-like length scale beyond which thermally excited undulations lead to a softened Young’s modulus, while the polymer nevertheless remains semiflexible. Both perturbative calculations and numerical Monte Carlo simulations suggest a qualitative change in several scaling properties of the buckling transition. The critical compressional strain for thermal buckling now increases with system size, in contrast to athermal buckling, where it decreases with system size. Renormalization group calculations confirm this picture, and also show that thermal buckling is controlled by a new fixed point with different critical exponents compared to classical Euler buckling. 

\end{abstract}

\maketitle

\section{Introduction}

Thermal fluctuations endow polymers with effective, scale-dependent elastic properties \cite{rubinstein2003polymer}, leading to dramatic consequences for both single molecule biophysics and nanoscience. \emph{Linear} polymers, such as single-stranded DNA, behave as a stiff rod when short, but when long, thermal fluctuations cause the polymer to wander as a random walk, allowing it to behave as a soft, entropic spring with a small, temperature dependent elastic modulus \cite{smith1996overstretching}. In contrast, two-dimensional \textit{sheet} polymers such as graphene exhibit thermal fluctuation induced stiffening, with a bending rigidity that is not only much larger than its estimated microscopic value, but also scales with system size \cite{blees2015graphene}.

Given these strongly scale dependent effects, a natural question is - what happens at a mechanical instability? The simplest instability in slender rods is Euler buckling \cite{landau2012}. That buckling can occur on small scales, where fluctuations dominate, is crucial for biological processes, such as cellular mechanics and force generation by the cytoskeleton, which is composed of linear, rod-like polymers such as microtubules and actin \cite{dogterom1997measurement, chaudhuri2007reversible, broedersz2014,brangwynne2006microtubules,lenz2012contractile}.  While previous work has investigated the effects of thermal fluctuations on the Euler buckling instability of linear polymers, much of it has been limited to inextensible polymers, where the contour length is fixed \cite{odijk1998microfibrillar,emanuel2007buckling,baczynski2007stretching,kierfeld2008semiflexible,kierfeld2010modelling,pilyugina2017buckling}. But polymers have a finite (albeit large) elastic stiffness that permits small stretching deformations. How these deformations geometrically couple to soft bending modes, thermal fluctuations, and external compression, remains poorly understood.

More recent work has begun exploring these effects for small fluctuations using standard perturbation theory \cite{bedi2015} or mean field analysis \cite{stuij2019}. However, studies of thermalized buckling in large polymerized sheets suggest that strong fluctuations can have a dramatic impact on the transition, for example, by modifying critical scaling exponents \cite{shankar2021,le2021thermal}. Can similar effects appear in the buckling  of fluctuating linear polymers?

In this paper, we address this question by considering a long semiflexible polymer compressed in an isometric ensemble (i.e., by fixing the endpoints of the polymer) at finite temperature. For a slender filament of rest length $L_0$, stretching modulus $Y$, and bending modulus $\kappa$, the persistence length at finite temperature $T$ is $\ell_p=\kappa/k_BT$ \cite{rubinstein2003polymer}, so filaments with $L_0\gg \ell_p$ are well-described as random walks. Instead, here we focus on the ``semiflexible'' regime, where $L_0\lesssim  \ell_p$ and an extensible, rod-like description is valid.
By combining renormalization group calculations and Monte Carlo simulations, we demonstrate that thermal fluctuations lead to an enhanced softening of the stretching modulus despite the polymer maintaining its semiflexible, rod-like elasticity. This leads to a larger critical compression threshold for buckling, consistent with previous perturbative results \cite{bedi2015}. Through a Ginzburg-like criterion we identify a thermal length scale  $
\ell_{\rm{th}}\sim (\kappa^2/Yk_BT)^{1/3}$ beyond which nonlinear fluctuation effects dominate and modify the critical scaling near the buckling transition. Notably, by estimating these length scales, we find that $\ell_{\rm{th}}\ll \ell_p$ by $2-3$ orders of magnitude for many systems of interest in biology and nanoscience (Table~\ref{tab:length_scales}), suggesting that non-classical scaling exponents should be observable in experiments. Our results shed new light on a finite temperature mechanical instability, with relevance for single molecule biophysics and nanomechanical metrology.

The paper is organized as follows. In Section \ref{sec:setup}, we set up the system and the elasticity theory of a linear polymer modeled as a thin extensible filament. We focus on the situation where the compression is implemented within a fixed strain mechanical ensemble, known as the isometric ensemble. (See \cite{shankar2021} for the distinction between the isometric and the more usual isotensional ensemble in two-dimensional buckling transitions.) In Section \ref{sec:euler}, we formally define the buckling transition and the exponents describing the mechanical scaling near the transition. Once we account for thermal fluctuations, we find that there is a scale-dependent softening of the renormalized 1D Young's modulus, arising because transverse displacements in the polymer result in stored length. This result also implies that the critical compression needed for thermal buckling is increased compared to athermal buckling. In Section \ref{sec:Ginzburg}, we use this calculation to estimate the softening of the Young's modulus using simple perturbation theory and define the thermal length $\ell_{\rm{th}}$.

In order to study the scaling properties near the thermal buckling transition, we employ both analytical and computational methods. In Section \ref{sec:rg}, we implement a one loop, fixed dimension momentum shell renormalization group procedure. Complementary analytical calculations are included in the appendices. Appendix \ref{app:free_energy} explains an alternative derivation of the free energy. Appendix \ref{app:eps} discusses the renormalization group calculation for an abstract model of a polymer with $D$ internal dimensions that allows a controlled expansion in $\varepsilon=4-D$. We also discuss a connection between the fixed strain (isometric) and fixed force (isotensional) mechanical ensembles. In Appendix \ref{app:saddlept}, we make use of a saddle point approximation to obtain an alternative estimate of the thermalized critical strain. Finally, in Section \ref{sec:mc}, we present Monte Carlo simulations of a discrete polymer model to test our analytical predictions, with some additional numerical analyses in Appendix \ref{app:binder_cumulant}, and conclude with a brief discussion in Section~\ref{sec:conclusions}. 

\section{Polymer elasticity in the isometric ensemble} \label{sec:setup}

\begin{figure*}
    \centering
    \includegraphics[width=1\linewidth]{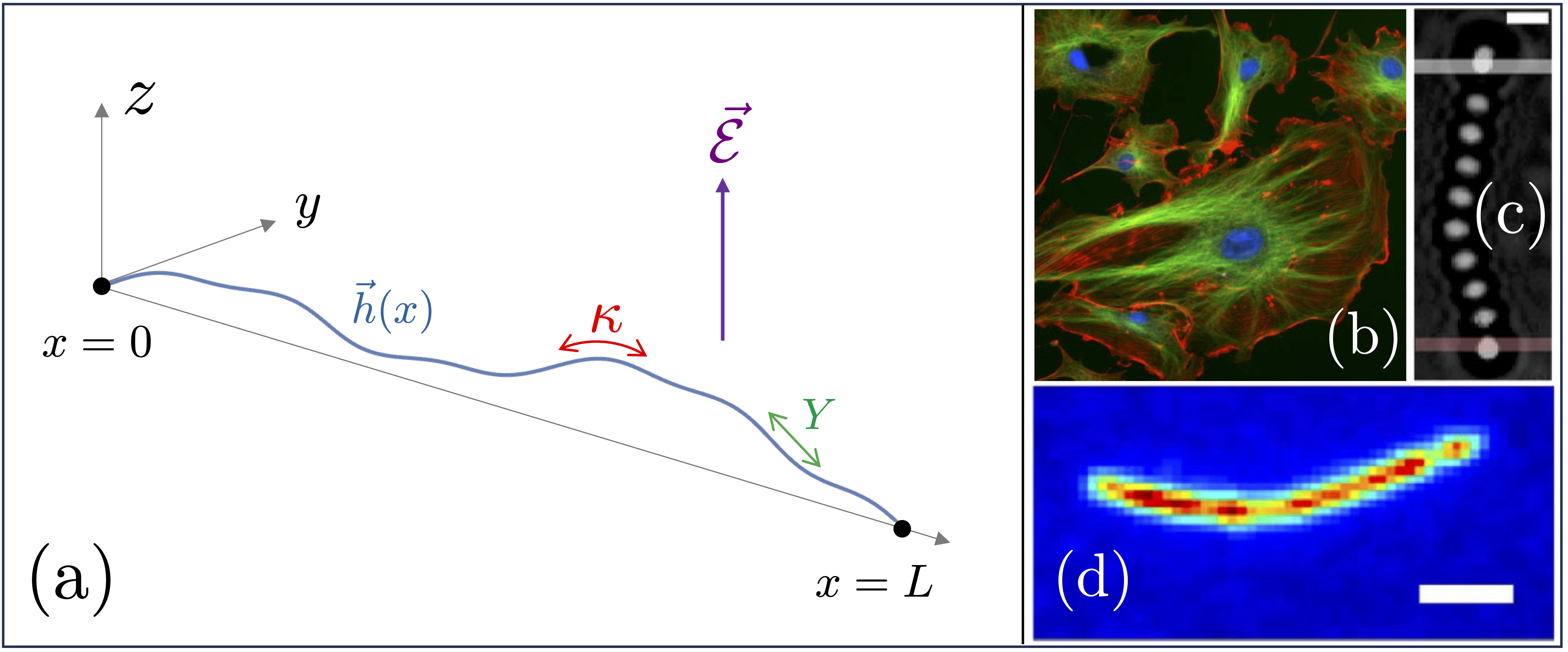}
    \caption{(a) Modeling a polymer in an isometric ensemble. The polymer has a zero temperature rest length $L_0$, and a fixed compressive strain $\epsilon$ is imposed by fixing the endpoints in place a distance $L=L_0(1-\epsilon)$ apart. Examples of physically relevant systems: (b) Fluorescence microscropy image of bovine pulmonary artery endothelial cells - microtubules (green), actin filaments (red), nuclei (blue). Source: example image from ImageJ (public domain) \cite{imagej} (c) Self-assembled colloidal chain from Ref. \cite{stuij2019} (scale bar: 3 $\mu$m) (d) Carbon nanotube from Ref. \cite{fakhri2009} (scale bar: 2 $\mu$m)}
    \label{fig:setup}
\end{figure*}

A semiflexible polymer can be modeled as a slender elastic rod that can both bend and stretch. We consider a rod of rest length $L_0$ (at zero temperature) to be composed of a homogeneous material with a three-dimensional (3D) Young's modulus $E$,with cross-sectional area $A$ and moment of inertia $I$ about the bending axis, which yields its bending rigidity to be $\kappa=EI$, and one-dimensional (1D) Young's modulus as $Y=EA$ \cite{landau2012}. In the isometric ensemble, the endpoints are separated by a fixed distance $L$, that we choose to lie along the $x$-axis, as depicted in Fig.~\ref{fig:setup}. For simplicity, we assume hinged boundary conditions, but we expect qualitatively similar results for other types of boundary conditions, such as clamped boundaries, as discussed in Section \ref{sec:euler}. Polymer deformations transverse to the $x$-axis are captured by a height field $\mathbf{h}(x)$, which in general, is a vector with $d_c$ (the codimension) number of components. The physically relevant scenarios of $d_c=1$ and $d_c=2$ correspond to the polymer being confined to a 2D plane and moving freely in 3D space, respectively. In general, transverse deformations imply the polymer is curved, with a contour length 
\begin{equation} \label{eq:contour}
    \tilde{L}=\int_0^L\sqrt{1+(\dd\mathbf{h}/\dd x)^2}\,\dd x.
\end{equation}

The elastic energy $F(\mathbf{h})$ of the polymer includes both bending ($F_b$) and stretching ($F_s$) contributions. The bending energy $F_b$ penalizes curvature and is given by 
\begin{equation} \label{eq:Fb}
F_b=\frac{\kappa}{2}\int_0^{\tilde{L}}\left(\frac{\dd\hat{\mathbf{t}}}{\dd s}\right)^2 \dd s\approx\frac{\kappa}{2}\int_0^L\left(\frac{\dd^2\mathbf{h}}{\dd x^2}\right)^2 \dd x,
\end{equation}
where $\hat{\mathbf{t}}=\left(\hat{\mathbf{x}}+\frac{\dd\mathbf{h}}{\dd x}\right)/\sqrt{1+(\dd\mathbf{h}/\dd x)^2}$ is the unit tangent and $\dd s=\sqrt{1+(\dd\mathbf{h}/\dd x)^2}\,\dd x$ is the infinitesimal arc length. The stretching energy $F_s$ is related to the change in length of the polymer and given by
\begin{equation}
\begin{split}
F_s&= \frac{1}{2}\frac{Y}{L_0}(\tilde{L}-L_0)^2\\
&=\frac{1}{2}\frac{Y}{L_0}[(\tilde{L}-L)+(L-L_0)]^2\\
&=\frac{1}{2}\frac{Y}{L_0}(\tilde{L}-L)^2-f(\tilde{L}-L)+\frac{1}{2}\frac{Y}{L_0}(L-L_0)^2.
\end{split}
\end{equation}
The last term in the third line is a constant and can be ignored, while we define the parameter $f\equiv Y\epsilon$, where the imposed compressional strain in the isometric ensemble is 
\begin{equation}
    \epsilon\equiv\frac{L_0-L}{L_0}.    
\end{equation}
According to our definition, $f>0$ corresponds to a compression, which is relevant for buckling. Let us note that while $f$ has dimensions of a force, it is actually the strain $\epsilon$ which is fixed in the isometric ensemble. As for the change in length $\tilde{L}-L$, a gradient expansion of Eq. (\ref{eq:contour}) gives
\begin{equation} \label{eq:stretch}
\tilde{L}-L\approx\int_0^L\left[\frac{1}{2}\left(\frac{\dd\mathbf{h}}{\dd x}\right)^2-\frac{1}{8}\left(\frac{\dd\mathbf{h}}{\dd x}\right)^4\right]\dd x,
\end{equation}
which leads to a stretching energy of the form
\begin{equation} \label{eq:Fs}
    \begin{split}
        F_s\approx-\frac{f}{2}\int_0^L\left(\frac{\dd\mathbf{h}}{\dd x}\right)^2\dd x&+\frac{f}{8}\int_0^L\left(\frac{\dd\mathbf{h}}{\dd x}\right)^4\dd x\\
        &+\frac{Y}{8L_0}\left[\int_0^L\left(\frac{\dd\mathbf{h}}{\dd x}\right)^2\dd x\right]^2.
    \end{split}
\end{equation}
Notably, the stretching energy includes a non-local quartic interaction that appears due to the fixed displacement boundary conditions imposed in the isometric ensemble \cite{bedi2015, shankar2021}. (See Appendix~\ref{app:free_energy} for an alternate derivation that integrates out elastic phonons along the rod axis.)

Aside from the bending and stretching energies, additional contributions to the free energy are possible, for instance, due to an applied external field $\bm{\mathcal{E}}$ (e.g., an electric field or gravity) that biases the filament to buckle in a particular direction. Upon combining all three contributions, we obtain
\begin{equation} \label{eq:F}
\begin{split}
F[\textbf{h}]=&\frac{\kappa}{2}\int_0^L \left(\frac{\dd^2\mathbf{h}}{\dd x^2}\right)^2 \dd x-\frac{f}{2}\int_0^L \left(\frac{\dd\mathbf{h}}{\dd x}\right)^2 \dd x\\
&+\frac{f}{8}\int_0^L\left(\frac{\dd\mathbf{h}}{\dd x}\right)^4\dd x\\
&+\frac{Y}{8L_0}\left[\int_0^L \left(\frac{\dd\mathbf{h}}{\dd x}\right)^2\dd x\right]^2-\bm{\mathcal{E}}\cdot\int_0^L \mathbf{h}(x)\,\dd x.
\end{split}
\end{equation}
In the gradient expansion of the free energy, we have truncated the bending energy at second order and the stretching energy at fourth order; higher order terms are argued to be small and irrelevant on large scales in Appendix \ref{app:eps} within a renormalization group framework. Here and below, we primarily focus on the isometric ensemble, where the endpoints of the polymer are fixed in space. On the other hand, the isotensional ensemble, where the force applied at the ends is fixed, is also a physically relevant scenario. In Appendix \ref{app:eps}, we also comment on the relationship between the two mechanical ensembles, but otherwise focus on the isometric ensemble in the main text.

\section{Euler buckling and scaling exponents} \label{sec:euler}

Before addressing the impact of thermal fluctuations, we review the Euler buckling instability and couch its scaling properties in a statistical mechanics language. This is easily seen by employing a mean field approximation on the nonlinear free energy in Eq.~(\ref{eq:F}) which yields the familiar classical results for zero temperature Euler buckling. Without loss of generality, we take the external field to point the $z$-direction: $\bm{\mathcal{E}}=\mathcal{E}\hat{\mathbf{z}}$. If we assume the endpoints of the polymer are pinned but free to rotate, an appropriate mean field ansatz for the transverse fluctuations $\mathbf{h}(x)$ is the lowest buckling mode $\mathbf{h}(x)=h_M\sin(\pi x/L)\hat{\mathbf{z}}$. Upon substituting this ansatz into the free energy in Eq. (\ref{eq:F}), we obtain a Landau-like potential
	\begin{equation} \label{eq:Landau}
    \begin{split}
        F_{\rm MF}(h_M)=&\frac{\pi^2Y}{4L}\left[\frac{\kappa}{Y}\cdot\left(\frac{\pi}{L}\right)^2-\epsilon\right]h_M^2\\
        &+\frac{(2+\epsilon)\pi^4Y}{64 L^3}\,h_M^4-\frac{2\mathcal{E}L}{\pi}\,h_M,
    \end{split}
	\end{equation}
which takes the form of a double well potential in the amplitude $h_M$, but with size-dependent coefficients \cite{hanakata2021thermal,stuij2019}. This unconventional size dependence arises because buckling is a mechanical instability of the mode with the smallest wavevector, which in this case is $\pi/L$. A key implication of the size-dependent coefficients is unconventional scaling relations between critical exponents, analogous to similar relations found for thermalized buckling of sheets \cite{shankar2021}. 

In the absence of an external field ($\mathcal{E}=0$), there is a critical compressive strain
\begin{equation} \label{eq:criticalpt}
    \epsilon_c=\frac{\kappa}{Y}\cdot\left(\frac{\pi}{L}\right)^2
\end{equation}
beyond which the polymer will buckle in a transverse direction, spontaneously breaking the up-down symmetry of the system in two dimensions and the rotational symmetry of the system in three dimensions. If we model the polymer as a solid cylindrical rod with a cross section of radius $r$, then the ratio of the bending and stretching moduli scales as $\kappa/Y\sim r^2$ \cite{landau2012}. Thus, the critical strain scales as $\epsilon_c\sim (r/L)^2$, approaching zero as the system size increases.

Upon minimizing the Landau potential in Eq. (\ref{eq:Landau}) with respect to $h_M$, we find that
\begin{equation}
|h_M|=
\begin{cases} 
\sqrt{\frac{8}{(2+\epsilon)\pi^2}}\,L(\epsilon-\epsilon_c)^{1/2} & \; \text{for $\epsilon>\epsilon_c$,} \\ 
0 & \; \text{for $\epsilon<\epsilon_c$.}
\end{cases}
\end{equation}
Within the mean field context, $h_M$ serves as an order parameter for the buckling transition, and since $h_M$ changes continuously across $\epsilon_c$, buckling is a supercritical bifurcation and can be viewed as a continuous phase transition. In later sections, we will take into account thermal fluctuations beyond mean field theory, in which case an appropriate order parameter is then the spatially-averaged transverse deformation $\left<\mathbf{h}\right>$:
\begin{equation}
\left<\mathbf{h}\right>\equiv\frac{1}{L}\int_0^L\mathbf{h}(x)\,\dd x.
\end{equation}
The critical exponent $\beta$ is defined by the power-law scaling of the order parameter in the buckled phase near the transition: $\left|\left<\mathbf{h}\right>\right|\propto |\epsilon-\epsilon_c|^\beta$. Thus, the mean field value is $\beta=1/2$.

The zero-field susceptibility $\chi$ measures the response of the order parameter to a small external field, and it diverges on both sides of the transition near the critical strain. The critical exponent $\gamma$, which is expected to be the same on both sides of the transition, captures the divergence:
\begin{equation}
\chi\equiv\left(\frac{\partial \left|\left<\mathbf{h}\right>\right|}{\partial \mathcal{E}}\right)_{\mathcal{E}=0}\propto |\epsilon-\epsilon_c|^{-\gamma}.
\end{equation}
Upon calculating $\chi$ using the Landau potential in Eq. (\ref{eq:Landau}), we obtain a mean field value of $\gamma=1$:
\begin{equation}
\chi=\left(\frac{\partial h_M}{\partial \mathcal{E}}\right)_{\mathcal{E}=0}=
\begin{cases} 
\frac{2}{\pi^3}\frac{L^2}{Y}(\epsilon-\epsilon_c)^{-1} & \; \text{for $\epsilon>\epsilon_c$,} \\ 
\frac{4}{\pi^3}\frac{L^2}{Y}(\epsilon_c-\epsilon)^{-1} & \; \text{for $\epsilon<\epsilon_c$.}
\end{cases}
\end{equation}

At precisely the critical point $\epsilon=\epsilon_c$, the order parameter responds nonlinearly to an applied external field $\mathcal{E}$, which defines the scaling exponent $\delta$ via $\left|\left<\mathbf{h}\right>\right|\propto \mathcal{E}^{1/\delta}$. Within mean field theory, we calculate the relation between the buckling amplitude $h_M$ and the external field $\mathcal{E}$ at the critical strain $\epsilon_c$:
\begin{equation}
h_M=\left(\frac{32}{(2+\epsilon_c)\pi}\right)^{1/3}\frac{L^{4/3}}{Y^{1/3}}\,\mathcal{E}^{1/3},
\end{equation}
and this implies a mean field value of $\delta=3$. 

We remark that our mean field results are derived using an ansatz whose choice is dictated by the boundary conditions on $\b{h}$. Although alternate boundary conditions would yield a Landau potential with different coefficients, for e.g.,  tangential clamping would require a buckling ansatz $\mathbf{h}(x)=\frac{h_M}{2}\left[1-\cos(2\pi x/L)\right]\hat{\mathbf{z}}$ which shifts the buckling threshold to $\epsilon_c=4\pi^2\kappa/(YL^2)$, these changes do not affect the critical exponents and qualitative behavior of the system near the transition. 

Finally, two additional critical exponents ($\eta$ and $\nu$)  characterize the spatial fluctuations of the polymer near the buckling transition. Right at the buckling transition, the Fourier transform $\mathbf{h}(x)=\frac{1}{L}\sum_q e^{iqx}\mathbf{h}_q$ of the $\mathbf{h}-\mathbf{h}$ correlation function scales as a power law, with an anomalous scaling exponent $\eta$:
\begin{equation}
\left<|\mathbf{h}_q|^2\right>\propto q^{-(4-\eta)}.
\end{equation}
Within a Gaussian/mean field description, where we assume the fluctuations are small enough to ignore the nonlinear terms in Eq. (\ref{eq:F}), we have a free energy 
\begin{equation} \label{eq:F0}
F_0[\textbf{h}]=\frac{\kappa}{2}\int_0^L \left(\frac{\dd^2\mathbf{h}}{\dd x^2}\right)^2 \dd x-\frac{f}{2}\int_0^L \left(\frac{\dd\mathbf{h}}{\dd x}\right)^2 \dd x.
\end{equation}
We tune to the critical point by imposing a compressional strain $\epsilon_c\sim 1/L^2$ that vanishes for a large system, so $f=Y\epsilon_c\to 0$. Thus, at the transition we obtain $\left<|\mathbf{h}_q|^2\right>\propto q^{-4}$ with $\eta=0$ in the Gaussian limit.

Away from the buckling transition, the imposed strain will lead to exponentially decaying tangent-tangent correlations on a scale set by the correlation length $\xi$: $\left<\hat{\mathbf{t}}(x)\cdot\hat{\mathbf{t}}(0)\right>\sim e^{-|x|/\xi}$, up to prefactors and subdominant terms. The critical exponent $\nu$ then captures the divergence of the correlation length as one approaches the critical point: $\xi\propto |\epsilon-\epsilon_c|^{-\nu}$. The tangent-tangent correlation function is, to lowest order in gradients given by
\begin{equation}
\left<\hat{\mathbf{t}}(x)\cdot\hat{\mathbf{t}}(0)\right>\approx 1-\frac{1}{2}\left<[\mathbf{h}'(x)-\mathbf{h}'(0)]^2\right>.
\end{equation}
From Eq. (\ref{eq:F0}), we have the propagator $\left<\mathbf{h}_q\cdot\mathbf{h}_k\right>=k_BTd_c L\delta_{q,-k}/(\kappa q^4-fq^2) $ which gives $\xi=\sqrt{\kappa/|Y\epsilon|}$ and thus $\nu=1/2$ in the Gaussian limit. While the propagator appears to have a pole at the nonzero $q=\sqrt{f/\kappa}$, this is an indicator of the buckling transition as an instability of the longest wavelength mode. Furthermore, we are interested in the vicinity of the buckling transition in Eq. (\ref{eq:criticalpt}), which corresponds to $f\to 0$. These mechanical scaling exponents are summarized in the ``Classical Euler'' column of Table \ref{tab:exponents}. In the following sections, we show how thermal fluctuations modify these results.

\begin{table*}
\caption{\label{tab:exponents} Summary of the differences in scaling exponents between classical Euler buckling and thermal isometric buckling.}
\begin{ruledtabular}
\def\arraystretch{1.5}
\begin{tabular}{ccccc}
 Exponent & Quantity & Scaling & Classical Euler \footnotemark[1]
& Thermal Isometric $(d_c=2)$ \footnotemark[2] \\ \hline
 $\beta$ & Order parameter $h$ & $h\sim L(\epsilon-\epsilon_c)^\beta$ & $1/2$ & $0.44$ \\
 $\gamma$ & Susceptibility $\chi$
 & $\chi\sim \frac{L^2}{Y}|\epsilon-\epsilon_c|^{-\gamma}$ & 1 & 0.89 \\
 $\delta$ & Nonlinear response at $\epsilon=\epsilon_c$ & $h\sim \frac{L^{4/3}}{Y^{1/3}}\mathcal{E}^{1/\delta}$ & $3$ & $3$\\
 $\eta$ & Anomalous scaling at $\epsilon=\epsilon_c$ & $\left<|\mathbf{h}(q)|^2\right>\sim q^{-(4-\eta)}$ & $0$ & $0$\\
 $\nu$ & Correlation length $\xi$ & $\xi\sim |\epsilon-\epsilon_c|^{-\nu}$ & $1/2$ & $0.44$\\
\end{tabular}
\end{ruledtabular}
\footnotetext[1]{Mean field values corresponding to Gaussian fixed point in the RG, applicable to systems $L<\ell_{\rm{th}}\sim (\kappa^2/Yk_BT)^{1/3}$, discussed in Section \ref{sec:euler}.}
\footnotetext[2]{Estimated from a non-trivial renormalization group fixed point, applicable to systems $\ell_{\rm{th}}<L<\ell_p\sim \kappa/k_BT$, discussed in Section \ref{sec:rg}}
\end{table*}

\section{Thermal fluctuations} \label{sec:thermal_fluctuations}

We now go beyond mean field theory and account for how thermal fluctuations lead to qualitative changes for an extensible, semiflexible polymer near its buckling transition. First, we find that thermal fluctuations lead to an effective softening of the Young's modulus, which is scale dependent at the buckling transition. Second, we find that the critical strain is delayed to a higher compressive strain, and the critical strain \textit{increases} with system size, in sharp contrast to the classical Euler buckling transition. This delay in the critical buckling threshold is consistent with earlier work on extensible rods \cite{bedi2015}. Finally, we find that there are new mechanical scaling exponents controlled by a nontrivial renormalization group fixed point, as described in the ``Thermal Isometric" column of Table \ref{tab:exponents}. These phenomena emerge from the nonlinear coupling between the imposed load, and both bending and stretching deformations. To deal with the nonlinear free energy in Eq. \ref{eq:F} systematically, we begin with a simple perturbation theory in Section \ref{sec:Ginzburg}. For large system sizes, perturbation theory breaks down and we implement a momentum shell renormalization group in Section \ref{sec:rg} to study the critical scaling behavior near thermal buckling.

\begin{figure}
    \centering
    \includegraphics[width=1\linewidth]{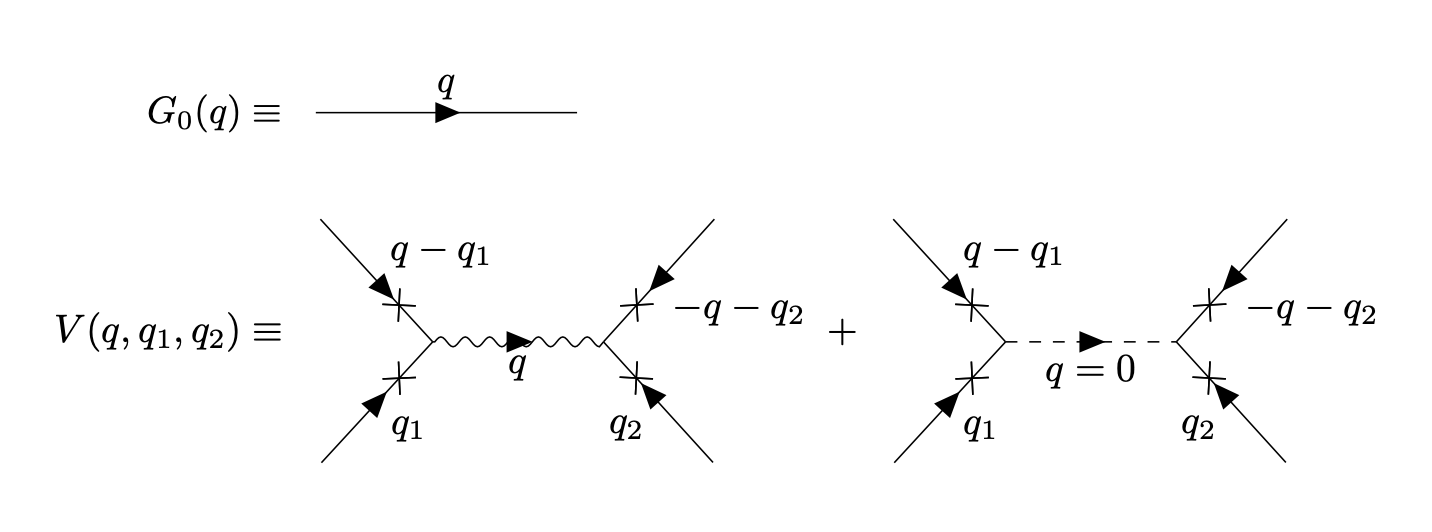}
    \caption{Diagrammatic notation for the bare propagator and quartic interactions in the transverse deformation field $\mathbf{h}$. The second, hyperlocal contribution to the quartic interactions is present even for $f=0$.}
    \label{fig:diagram_setup}
\end{figure}

It is useful to set up a diagrammatic representation of the nonlinear free energy. In Fourier space $\mathbf{h}(x)=\frac{1}{L}\sum_q e^{iqx}\mathbf{h}_q$, the free energy with zero external field from Eq. (\ref{eq:F}) is
\begin{equation}
\begin{split}
F[\mathbf{h}_q]&=\frac{1}{2}\int_q(\kappa q^4-fq^2)|\mathbf{h}_q|^2\\
&+\int_q\int_{q_1}\int_{q_2}V(q,q_1,q_2)(\mathbf{h}_{q-q_1}\cdot\mathbf{h}_{q_1})(\mathbf{h}_{-q-q_2}\cdot\mathbf{h}_{q_2}).\\
\end{split}
\end{equation}
The bare propagator $G_0(q)=k_BT/(\kappa q^4-fq^2)$ and the nonlinear interactions
\begin{equation} \label{eq:vertex}
\begin{split}
V(q,q_1,q_2)=\frac{f}{8}&\cdot (q-q_1)q_1q_2(-q-q_2)\\
&+\frac{Y}{8L_0}\cdot 2\pi\delta(q)\cdot q_1^2q_2^2
\end{split}
\end{equation}
are represented diagrammatically using the line and vertices  depicted in Figure \ref{fig:diagram_setup}. 
In Eq. (\ref{eq:vertex}), there are two nonlinear quartic interactions that are encapsulated in $V(q,q_1,q_2)$, and they are represented using a wavy line for the $f$ term and a dashed line for the $Y$ term. In the diagrams, the arrows indicate momenta, while the cross marks indicate derivatives. Note that the latter $Y$ interaction is hyperlocal in Fourier space, as is typical for systems with constraints (in our case, the endpoints being fixed) \cite{rudnick1974,bedi2015,shankar2021}. In the following subsections, we will analyze the nonlinear free energy analytically, starting with simple perturbation theory before implementing a momentum shell renormalization group. 

\subsection{Perturbation theory and a Ginzburg criterion} \label{sec:Ginzburg}

\begin{figure}
    \centering
    \includegraphics[width=1\linewidth]{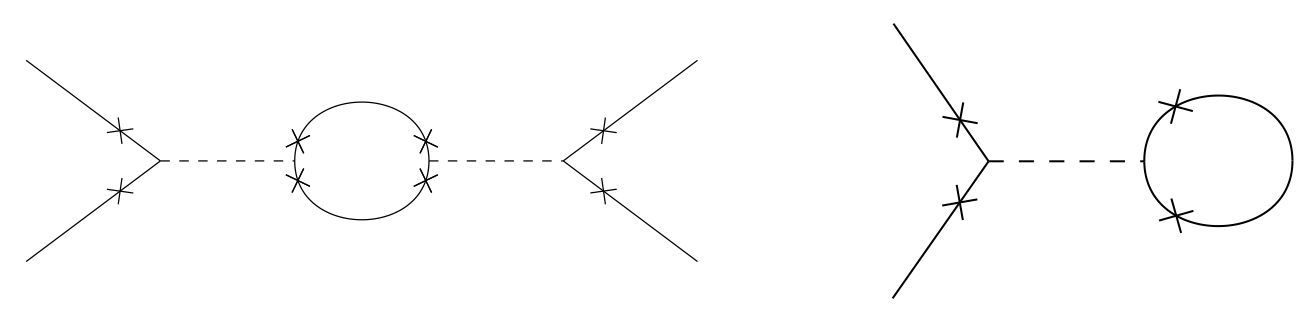}
    \caption{One loop corrections to the vertex and the propagator at the critical buckling transition $f_c\to 0$}
    \label{fig:diagram_perturbation}
\end{figure}

Thermal fluctuations lead to an effective softening in the Young's modulus of an extensible polymer. At finite temperature, the polymer will experience transverse undulations, which create stored length. This makes the polymer effectively softer, as energetically cheaper bending deformations in these regions couple to the applied load, making the polymer easier to stretch or compress. This intuitive picture is captured within a simple perturbative calculation of the one loop correction to the Young's modulus, as depicted in Figure \ref{fig:diagram_perturbation}. To evaluate the loop integral, we impose cutoffs set by the system size $L$ and a microscopic length scale $a$ that is comparable to the monomer spacing or rod diameter. 
In particular, we are interested in the perturbative correction when the system is tuned to its critical strain $\epsilon_c\sim 1/L^2$, so $f_c=Y\epsilon_c\to 0$ for large $L$. In this case, only the second term in Eq. (\ref{eq:vertex}) matters, and we find that the renormalized Young's modulus $Y_R$ has a perturbative correction that scales with the cube of the system size $L$:
\begin{equation}
\begin{split}
    \frac{Y_R}{Y}&= 1-4d_c\cdot\frac{Yk_BT}{8}\cdot 2\int_{\pi/L}^{\pi/a}\frac{\dd k}{2\pi}\frac{1}{\kappa^2 k^4} \\
    &\approx 1 - \frac{d_c Yk_BT}{6\pi^4\kappa^2}L^3.
\end{split}
\end{equation}
This expansion immediately highlights its own demise as fluctuations, no matter how small a temperature, inevitably drive $Y_R$ towards zero for a large enough system. By using a Ginzburg criterion, we identify a thermal length scale $\ell_{\rm{th}}$ beyond which the fluctuation induced softening of $Y$ is comparable to its bare value. This gives a thermal Ginzburg length, 
\begin{equation} \label{eq:l_th}
\ell_{\rm{th}}\equiv\left(\frac{6\pi^4}{d_c}\cdot\frac{\kappa ^2}{Yk_BT}\right)^{1/3}.
\end{equation}
The scale dependent correction means that for sufficiently large systems $L>\ell_{\rm{th}}$, perturbation theory breaks down, and this hints at a change in scaling behavior for finite temperature buckling when compared to classical, athermal buckling. For extensible polymers under \textit{tension}, a similar length scale has been used to estimate a crossover between an entropy-dominated and elasticity-dominated deformation regime \cite{odijk1995stiff}, but its consequences for buckling seem not to have been studied before.

For a polymer that can both bend and stretch, thermal fluctuations therefore lead to the emergence of two important length scales. The first length scale is the thermal Ginzburg length $\ell_{\rm{th}}$ from Eq. (\ref{eq:l_th}), which gives the length scale above which thermal fluctuations significantly soften the effective Young's modulus. We are interested in polymers with size $L>\ell_{\rm{th}}$. The second length scale is the familiar persistence length \cite{de1979scaling}
\begin{equation}
    \ell_p\equiv \frac{\kappa}{k_BT}\;,
\end{equation}
beyond which the polymer conformations are dominated primarily by their entropy. If we wish to compress the polymer and observe thermal buckling, we need to restrict ourselves to the semiflexible regime $L<\ell_p$ so that the polymer maintains its rod-like elasticity and a roughly straight phase in the absence of compression. 

To study such polymers that remain semiflexible, yet experience renormalization of their effective elastic properties, we focus on an intermediate range of system sizes $\ell_{\rm{th}}<L<\ell_p$. But what is the typical separation of length scales between $\ell_{\rm{th}}$ and $\ell_p$? In general, the thermal length scale $\ell_{\rm{th}}$ is shorter than the persistence length $\ell_p$. This is easily seen by noting that $\kappa\sim Y r^2$ \cite{landau2012} for a slender, solid rod of cross-sectional radius $r$, so $\ell_{\rm{th}}/\ell_p\sim (r/\ell_p)^{2/3}\ll 1$ as $r\ll \ell_p$ for a thin, stiff filament. This suggests there is a range of length scales between $\ell_{\rm{th}}$ and $\ell_p$ where thermal fluctuations can noticeably impact the scaling exponents near the buckling transition in semiflexible polymers. In Table \ref{tab:length_scales}, we estimate $\ell_{\rm{th}}$ and $\ell_p$ for experimental model polymers including microtubules, F-actin, carbon nanotubes, etc., \cite{broedersz2014, fakhri2009, stuij2019} and find a large range for $\ell_p/\ell_{\rm{th}}\sim 10^2-10^3$ suggesting an experimentally accessible range of system sizes with $\ell_{\rm{th}}< L<\ell_p$.
\begin{table}
\caption{\label{tab:length_scales}
Key length scales for extensible, semiflexible polymers at finite temperature. For simplicity, the polymers are assumed to be confined to a plane, corresponding to codimension $d_c=1$. Estimates are obtained using values from \textsuperscript{a} Ref. \cite{broedersz2014}, \textsuperscript{b} Ref. \cite{fakhri2009}, and \textsuperscript{c} Ref. \cite{stuij2019}.}
\begin{ruledtabular}
\def\arraystretch{1.5}
\begin{tabular}{cccc}
System & $\ell_{\rm{th}}\equiv(\frac{6\pi^4}{d_c}\cdot\frac{\kappa}{Y}\cdot \ell_p)^{1/3}$ & $\ell_p\equiv\frac{\kappa}{k_BT}$ & $\ell_p/\ell_{\rm{th}}$\\
\hline
Microtubule \footnotemark[1] & 4 $\mu$m & 3 mm & $\sim 750$ \\
F-actin \footnotemark[1] & 0.3 $\mu$m & 17 $\mu$m & $\sim 57$ \\
DNA \footnotemark[1] & 19 nm & 50 nm & $\sim 2.6$ \\
Carbon nanotube \footnotemark[2] & 0.2 $\mu$m & 100 $\mu$m & $\sim 500$ \\
Colloidal chain \footnotemark[3] & 100 $\mu$m & 2.9 mm & $\sim 29$ \\
\end{tabular}
\end{ruledtabular}
\end{table}

Finally, if thermal fluctuations lead to an effective softening of the Young's modulus, then one would also expect thermal effects to shift the critical compressive strain to a larger value compared to classical, athermal buckling. Intuitively, buckling occurs when it becomes energetically favorable to bend in a transverse direction rather than to keep compressing in a configuration that is straight on average. If the Young's modulus is effectively reduced, then one can compress the polymer with a larger strain before the buckling tradeoff to bending energy occurs. Within perturbation theory, we can estimate the shift in the critical strain by computing the one loop correction to the propagator, as depicted in Figure \ref{fig:diagram_perturbation}. The buckling transition at finite temperature is delayed from the zero temperature critical strain $\epsilon_c(T=0)$ in Eq. (\ref{eq:criticalpt}) to $\epsilon_c(T)=\epsilon(T=0)+\Delta\epsilon$, where 
\begin{equation} \label{eq:perturb_shift}
    \Delta \epsilon=d_ck_BT\int_{\pi/L}^{\pi/a}\frac{\dd k}{2\pi}\frac{1}{\kappa k^2}\approx\frac{d_ck_BTL}{2\pi^2\kappa}=\frac{d_c}{2\pi^2}\frac{L}{\ell_p}.
\end{equation}
The form of $\Delta \epsilon\sim d_c L/\ell_p$ is consistent with the value computed in \cite{bedi2015}. Notably, this diverging correction means that the finite temperature buckling threshold \textit{increases} with system size $L$, which is the opposite of trend for classical buckling, where the buckling threshold decreases with system size as $L^{-2}$. In Appendix \ref{app:saddlept}, the shift in the critical strain is also computed within an alternative saddle point approximation, with similar size-dependent behavior in the regime $\ell_{\rm{th}}<L<\ell_p$. In the following section, we will investigate other novel scaling behaviors associated with thermalized buckling of extensible polymers. 

\subsection{Momentum shell renormalization group} \label{sec:rg}

In the previous section, we saw how for polymers with a size $L>\ell_{\rm{th}}$, nonlinearities in the elastic free energy led to both a dramatic softening in the Young's modulus and a change in the scaling of the critical strain. In this section, we go beyond simple perturbation theory and implement a standard momentum-shell renormalization group (RG) procedure \cite{kardar2007, nelson2025renormalization}, which handles the infrared divergences in perturbation theory while revealing the scaling behavior near the thermal buckling transition. Here, for simplicity, we present a fixed dimension one-loop approximation, and leave a more controlled $\varepsilon=4-D$ expansion of a generalized, abstract polymer model to Appendix \ref{app:eps}. The two approaches yield the same qualitative results, though they differ quantitatively. The form of the elastic energy from Eq. (\ref{eq:F}) is 
\begin{equation} \label{eq:F_rg}
\begin{split}
F[\mathbf{h}]=\frac{\kappa}{2}&\int_0^L \left(\frac{\dd^2\mathbf{h}}{\dd x^2}\right)^2 \dd x-\frac{f}{2}\int_0^L \left(\frac{\dd\mathbf{h}}{\dd x}\right)^2 \dd x\\
&+\frac{g}{8}\int_0^L\left(\frac{\dd\mathbf{h}}{\dd x}\right)^4\dd x+\frac{y}{8L}\left[\int_0^L \left(\frac{\dd\mathbf{h}}{\dd x}\right)^2\dd x\right]^2,
\end{split}
\end{equation}
where we have introduced the coupling constants $g$ and $y$ for the two quartic nonlinearities. The coupling $y$ is related to the Young's modulus, $y=Y(1-\epsilon)$, and is introduced for notational simplicity, whereas $g=f$ at the microscopic level. We use a separate coupling for $g$ because upon renormalization, $f$ and $g$ do not necessarily evolve in the same fashion, as the isometric ensemble breaks the rotational symmetry of the system at the endpoints. 

\begin{figure*}
    \centering
    \includegraphics[width=1\linewidth]{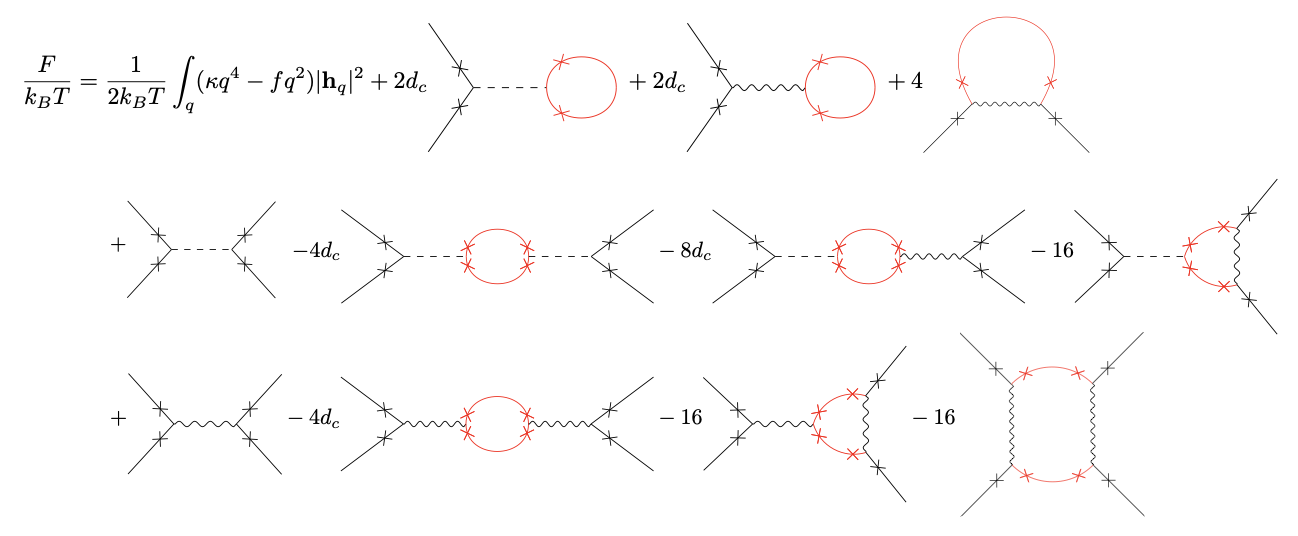}
    \caption{Momentum shell RG to one loop. The red lines indicate fast modes in the momentum shell to be integrated out. The first line shows the corrections to the propagator, the second line shows the corrections to the $y$ vertex, and the third line shows the corrections to the $g$ vertex.}
    \label{fig:diagram_rg}
    \end{figure*}

The system has an ultraviolet (UV) cutoff $\Lambda\sim 1/a$, where $a$ is a microscopic length scale such as the monomer spacing or rod diameter. To implement the momentum-shell renormalization group, we integrate over modes in a shell $\Lambda/b<|\b{q}|<\Lambda$ with $b\gtrsim 1$, and decompose the transverse field $\mathbf{h}(x)=(1/L)\sum_q \mathbf{h}_q e^{iqx}$ into fast ($\mathbf{h}_q^>$) and slow ($\mathbf{h}_q^<$) modes in Fourier space: $\mathbf{h}_q=\theta(\Lambda/b-q)\mathbf{h}_q^<+\theta(q-\Lambda/b)\mathbf{h}_q^>$, where $\theta$ is the Heaviside step function. Fast modes are represented diagrammatically in red, while slow modes are shown in  black in Fig. \ref{fig:diagram_rg}. Upon integrating out the fast modes in the momentum shell, we avoid any infrared divergences, and obtain finite renormalization contributions for the coupling constants $\kappa, f,y$, and $g$. The one-loop contributions are shown in Fig. \ref{fig:diagram_rg}. 
As a final step of the RG procedure, we rescale lengths and fields: $x=bx'$ and $\mathbf{h}(x)=b^\zeta\mathbf{h}'(x')$. Upon setting $l=\ln b\ll 1$, we obtain renormalization group  recursion relations:
\begin{equation} \label{eq:RG_recursion}
	\begin{split}
	\frac{\dd\kappa}{\dd l}&=(2\zeta-3)\kappa,\\
	\frac{\dd f}{\dd l}&=(2\zeta-1)f-\frac{d_c y+(d_c+2)g}{2\pi}\cdot \frac{\Lambda k_BT}{\kappa \Lambda^2-f},\\
	\frac{\dd y}{\dd l}&=(4\zeta-3)y-y\cdot \frac{d_cy+2(d_c+2)g}{2\pi}\cdot \frac{\Lambda k_BT}{(\kappa \Lambda^2-f)^2},\\
	\frac{\dd g}{\dd l}&=(4\zeta-3)g-\frac{(d_c+8)g^2}{2\pi}\cdot \frac{\Lambda k_BT}{(\kappa \Lambda^2-f)^2}.
	\end{split}
\end{equation}
The recursion relation for the bending rigidity $\kappa$ has no fluctuation correction, in contrast to thermally fluctuating sheets \cite{nelson1987fluctuations,le1992self,aronovitz1988fluctuations,nelson2004,le2018anomalous}. In the case of sheets, thermal fluctuations generically excite deformations with local Gaussian curvature, causing the sheet to effectively stiffen to bending on large scales. This geometric effect is absent in linear polymers, as a result we would indeed not expect renormalization of the  bending rigidity. The recursion relation for the effective force $f$ has a correction which makes the force less compressive, and the recursion relation for the Young's modulus ($\propto y$) displays a softening correction. The intuition for both these terms is similar: transverse fluctuations result in stored length, which makes it easier to stretch or compress the polymer. For free boundary conditions, this stored length means the end-to-end distance will prefer to be shorter than the natural length of the polymer. But in the isometric ensemble, the endpoints are fixed, so at zero imposed strain, the clamped boundaries will oppose the shrinkage and lead to a spontaneously generated tension in the polymer. 

In the isometric ensemble, rotational symmetry is broken by  fixing the endpoints of the polymer. At the microscopic level, $g=f$ in Eq. (\ref{eq:F_rg}), as both originate from the same bulk stretching term in Eq. (\ref{eq:stretch}). However, $f$ and $g$ renormalize differently, as reflected by their distinct recursion relations in Eq. (\ref{eq:RG_recursion}). This implies that while rotational symmetry was originally broken only at the boundaries, thermal fluctuations cause it to be broken in the bulk as well. Hence, the situation here is in contrast with the thermal isometric buckling of \textit{sheets} in \cite{shankar2021}, and is more similar to the buckling transition in \cite{le2021thermal}, where an external field is used to break the rotational symmetry of a thermalized membrane in the bulk. 

\begin{figure*}
	\centering
	   \includegraphics[width=0.9\linewidth]{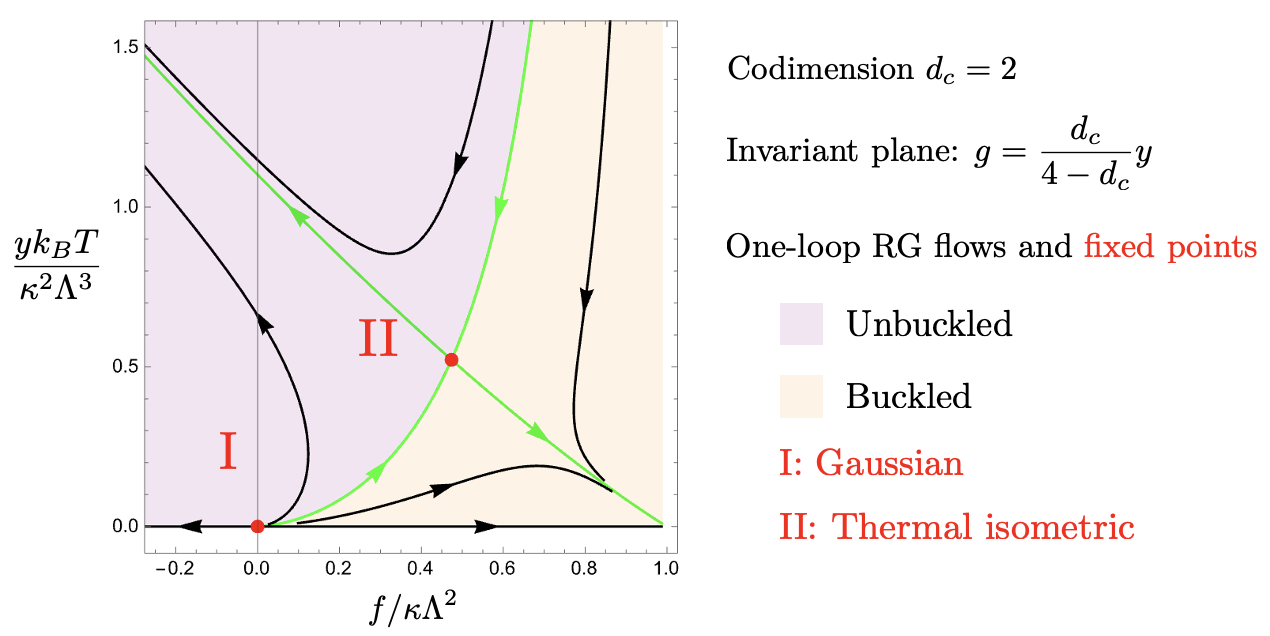} 
	   \caption{One loop RG flows within an invariant plane $g=\frac{d_c}{4-d_c}y$ for codimension $d_c=2$. This plane is an attractive subspace within the three dimensional parameter space of RG flows. There are two fixed points: I corresponds to classical Euler buckling, while II corresponds to thermal buckling in an isometric ensemble. The incoming part of the green separatrix connecting I and II marks the boundary between the buckled (orange) and unbuckled (purple) phases. Since the phase boundary curves rightward away from the vertical axis, the critical buckling compression is increased for finite temperature.}
    \label{fig:rg_flows}
	\end{figure*}

Without loss of generality, we pick a renormalization group rescaling factor $\zeta=3/2$, which allows us to set $d\kappa/dl=0$, and  analyze the renormalization group flows in the parameter space $(f,y,g)$. There are four fixed points, one of which is the Gaussian fixed point $(f^*,y^*,g^*)=(0,0,0)$ that describes the physics of classical, zero temperature Euler buckling. For codimension $d_c<4$, there is a novel interacting fixed point which describes the physics of \textit{thermal isometric} buckling:
\begin{equation}
\begin{split}
&\left(\frac{f^*}{\kappa\Lambda^2},\frac{y^*k_BT}{\kappa^2\Lambda^3},\frac{g^*k_BT}{\kappa^2\Lambda^3}\right)\\
&=\left(\frac{9}{d_c+17},\frac{6\pi(4-d_c)(d_c+8)}{d_c(d_c+17)^2},\frac{6\pi(d_c+8)}{(d_c+17)^2}\right).
\end{split}
\end{equation}
Notably, within the three-dimensional parameter space, there is an invariant plane of flows $g=[d_c/(4-d_c)]y$ which is an attractive subspace and contains both the Gaussian and thermal isometric fixed points. Within this invariant plane, the recursion relations take the simplified form
\begin{equation}
	\begin{split}
	\frac{\dd f}{\dd l}&=2f-\frac{3d_cy}{\pi(4-d_c)}\cdot \frac{\Lambda k_BT}{\kappa \Lambda^2-f},\\
	\frac{\dd y}{\dd l}&=3y-\frac{d_c(d_c+8)y^2}{2\pi(4-d_c)}\cdot \frac{\Lambda k_BT}{(\kappa \Lambda^2-f)^2},
	\end{split}
\end{equation}
and the resulting renormalization group flows are plotted in Figure \ref{fig:rg_flows}.
To determine critical exponents, we linearize the RG flow equations about the thermal isometric fixed point, and diagonalize the Jacobian matrix to obtain the positive (unstable) eigenvalue $\lambda>0$ which gives the critical exponent $\nu$ as
\begin{equation}
\begin{split}
\nu&=1/\lambda=\frac{2(d_c+8)}{\sqrt{3940+d_c(25d_c+652)}-(d_c+26)}\\&\approx \begin{cases} 0.440 \quad &\text{for } d_c=1, \\ 0.443 \quad &\text{for } d_c=2. \end{cases}
\end{split}
\end{equation}
Importantly, the value of $\nu$ differs from its mean field value of $1/2$, so thermal buckling of a linear polymer in the isometric ensemble leads to different exponents from classical Euler buckling. We can also estimate the anomalous scaling exponent $\eta$ at the thermal isometric fixed point. From the definition of the renormalized bending rigidity $\kappa_R(q)\sim q^{-\eta}$, we have $\left<h(x)^2\right>=\int_{\pi/L}^{\pi/a}\frac{\dd q}{\kappa_R(q)q^4}\sim L^{3-\eta}$. Since we work with the rescaling factor $\zeta=3/2$, at the thermal isometric fixed point, we compute $\left<h(x)^2\right>=\int_{\pi/L}^{\pi/a}\frac{\dd q}{\kappa q^4}\sim L^3$, which tells us that $\eta=0$. This is consistent with the picture that there is no thermal renormalization of the effective bending rigidity. Now that we have found the values of $\nu$ and $\eta$, scaling relations determine the remaining critical exponents. In the context of thermalized sheet polymers buckled in an isometric ensemble \cite{shankar2021}, the following scaling relations have been  derived:
\begin{equation}
\beta=\nu\left(1-\frac{\eta}{2}\right), \quad \gamma=2\beta, \quad \delta=3.
\end{equation}
While it is unclear if these relations hold for linear polymers (given the stronger impact of rotational symmetry breaking), a controlled RG calculation using an $\varepsilon$-expansion of a generalized version of our polymer problem (Appendix \ref{app:eps}) demonstrates that these relations indeed remain true to leading order in $\varepsilon$.

Notably, the exponents $\nu, \beta$, and $\gamma$ at the thermal isometric fixed point are distinct from those for Euler buckling. The results for the critical exponents are summarized in the ``Thermal Isometric" column of Table \ref{tab:exponents}. In addition, we can estimate the scale-dependent softening of the effective Young's modulus $Y_R$ by computing the irrelevant eigenvalue $\lambda_{\rm irr}\approx -5.06$ at the thermal isometric fixed point, which suggests $Y_R/Y\sim (L/\ell_{\rm{th}})^{-5.06}$. Within this one loop approximation, while the qualitative predictions for the scaling behavior are expected to be true, the exact values of the scaling exponents may not be accurate, similar to one loop renormalization group treatments of free-standing sheet polymers \cite{nelson2025renormalization}.

\begin{figure}
    \centering
    \includegraphics[width=1\linewidth]{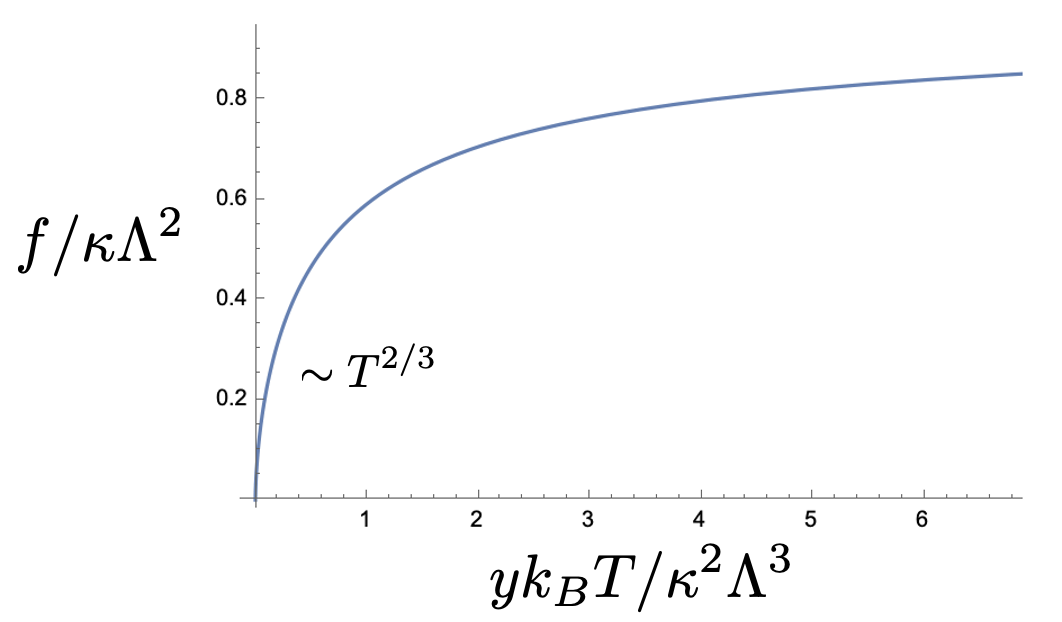}
    \caption{An additional thermal shift in the critical strain in the isometric ensemble within our one loop RG calculation. It is size independent, and scales as $\delta \epsilon\sim \ell_{\rm{th}}/\ell_p\sim T^{2/3}$.}
    \label{fig:rg_shift}
\end{figure}

In addition to the scaling exponents, the renormalization group analysis also provides an estimate for the shift in the buckling transition due to nonlinear fluctuation corrections. In Figure \ref{fig:rg_flows}, the green incoming separatrix connecting the Gaussian and thermal isometric fixed points marks the boundary between unbuckled and buckled phases. For any initial condition to the left of the separatrix, the renormalization group leads to a flow to large, negative $f$, which indicates a tensile strain at long wavelengths and a flat phase. Meanwhile, for any initial condition to the right of the separatrix, the renormalization group leads to a flow to large, positive $f$, which indicates a compressional strain and a buckled phase. By determining the shape of the separatrix, we can determine the fluctuation induced shift $\delta\epsilon$ for the thermal buckling transition in the isometric ensemble, as depicted in Figure \ref{fig:rg_shift}.
For low temperatures, we find that 
\begin{equation}
\frac{Y\delta\epsilon}{\kappa\Lambda^2}=c\left(\frac{Yk_BT}{\kappa^2\Lambda^3}\right)^{2/3} \Longrightarrow \delta\epsilon=c\left(\frac{d_c}{6\pi^4}\right)^{1/3}\frac{\ell_{\rm{th}}}{\ell_p},
\end{equation}
where the prefactor is estimated numerically to be $c\approx 1.46$. While the renormalization group method captures fluctuation effects beyond that of the perturbation theory calculation in Section \ref{sec:Ginzburg}, it does not capture finite system size effects, as the momentum-shell method integrates only over a continuum of high-momenta/short-wavelength modes. To estimate the critical strain due to both effects, we use
\begin{equation}
\epsilon_c(T) \approx \epsilon_c(T=0)+\Delta \epsilon+\delta \epsilon,
\end{equation}
where $\Delta \epsilon$ is from Eq. (\ref{eq:perturb_shift}). Note however, that $\ell_{\rm{th}}\ll \ell_p$ for our systems of interest, and so this new, additional fluctuation-induced shift $\delta \epsilon$ is small. 

\section{Comparison to simulations} \label{sec:mc}

To test some of our analytical predictions, we perform Markov chain Monte Carlo simulations  of an extensible, semiflexible polymer in an isometric ensemble and numerically probe its thermalized buckling. For simplicity, we confine the polymer to a plane, corresponding to codimension $d_c=1$. These simulations check the anomalous scaling of the critical strain, which is predicted to increase with system size, as well as the size-dependent softening of the effective Young’s modulus. Within conventional finite size scaling approaches, the remaining critical exponents are difficult to extract, since the restriction that the system size must be less than the persistence length means there is not a simple thermodynamic limit (see the end of this section as well as Appendix \ref{app:binder_cumulant} for additional discussion). To set up the system, we discretize the polymer into $N$ identical springs, each with rest length $l_0$ and spring constant $k_s=Y/l_0$. Deviations in the angle between adjacent springs are energetically penalized by a bending modulus $k_b=\kappa/l_0$. The end points of the polymer are fixed at a desired separation to impose compression in the isometric ensemble. While displacements (both longitudinal and transverse) are fixed to be zero at the end points, the boundary conditions are hinged, so the tangents at ends are free to rotate. The discretized elastic energy is thus
\begin{equation} \label{eq:discrete_energy}
    \begin{split}
        E&=k_b\sum_{i=1}^{N-1}(1-\cos\theta_i)+\frac{1}{2}k_s\sum_{i=1}^N(|\mathbf{r}_{i}-\mathbf{r}_{i-1}|-l_0)^2,\\
        &\cos\theta_i=\frac{(\mathbf{r}_{i+1}-\mathbf{r}_{i})\cdot(\mathbf{r}_{i}-\mathbf{r}_{i-1})}{|\mathbf{r}_{i+1}-\mathbf{r}_{i}||\mathbf{r}_{i}-\mathbf{r}_{i-1}|}.
    \end{split}
\end{equation}
In the limit where $N\to \infty$ and $l_0\to 0$, but with $Nl_0=L_0$ fixed, one can recover the continuum elastic energy, analogous to when one discretizes a membrane using a triangulated mesh \cite{seung1988defects}. Note we did not include the final biasing term in Eq. (\ref{eq:F}), corresponding to an applied external field, in these simulations. 

Thermal fluctuations are incorporated within a Markov chain Monte Carlo (MCMC) procedure, implemented using a Metropolis algorithm  \cite{binder1992}. To speed up thermal equilibration, two types of trial moves (local and wave) are used. Local trial moves attempt a small random displacements of a randomly chosen interior node, so $(x_i,y_i)\to(x_i+\xi_1,y_i+\xi_2)$ for $i=1,\cdots,N-1$, where $\xi_{1,2}\in[-\Delta,\Delta]$ are independent, uniformly distributed random variables. Wave trial moves are nonlocal and attempt to shift the overall profile of the polymer by a random Fourier mode of the system, i.e.,  $y_i\to y_i+M_q\sin(qx_i)$, where $M_q\in[-\Delta_q, \Delta_q]$ is a uniformly distributed random variable and $q=n\pi/L$ for $n=1,\cdots, \lfloor N/10\rfloor$. Combining both local and Fourier (wave) updates in MCMC has previously been used in the study of inextensible polymers and thermalized graphene \cite{baczynski2007stretching, kierfeld2008semiflexible, los2009scaling, granato2025thermal}. For both kinds of trial moves, a standard Metropolis acceptance protocol is used to ensure detailed balance, where a trial move from state $i$ to state $j$ is accepted with probability $1$ if $E_j\le E_i$ and with probability $e^{-(E_j-E_i)/k_BT}$ if $E_j>E_i$. The values of $\Delta$ and the $\Delta_q$'s were chosen such that the acceptance rate of each type was roughly $50\%$.

It is useful to non-dimensionalize the problem by writing $\overline{E}\equiv E/k_BT$ and $\mathbf{s}_i=\mathbf{r}_i/l_0$, so that Eq. (\ref{eq:discrete_energy}) becomes
\begin{equation}
    \begin{split}
        \overline{E}&=s_p\left[\sum_{i=1}^{N-1}(1-\cos\theta_i)+\frac{1}{2}\omega\sum_{i=1}^N(|\mathbf{s}_{i}-\mathbf{s}_{i-1}|-1)^2\right],\\
        &\cos\theta_i=\frac{(\mathbf{s}_{i+1}-\mathbf{s}_{i})\cdot(\mathbf{s}_{i}-\mathbf{s}_{i-1})}{|\mathbf{s}_{i+1}-\mathbf{s}_{i}||\mathbf{s}_{i}-\mathbf{s}_{i-1}|},
    \end{split}
\end{equation}
with the dimensionless persistence length $s_p\equiv \kappa/(l_0k_BT)$ and an analogue of a local F{\"o}ppl von Karman number $\omega\equiv Yl_0^2/\kappa$ that measures the relative importance of stretching to bending on the small scale. We set $s_p=\omega=1000$, which gives the maximum size $N_{\rm{max}}=\ell_p/l_0=10^3$ for the elastic rod-like description to hold, and a minimum size $N_{\rm{min}}=\ell_{\rm{th}}/l_0\approx 10$, for thermal fluctuations to be important.

\begin{figure*}
    \centering
    \includegraphics[width=0.93\linewidth]{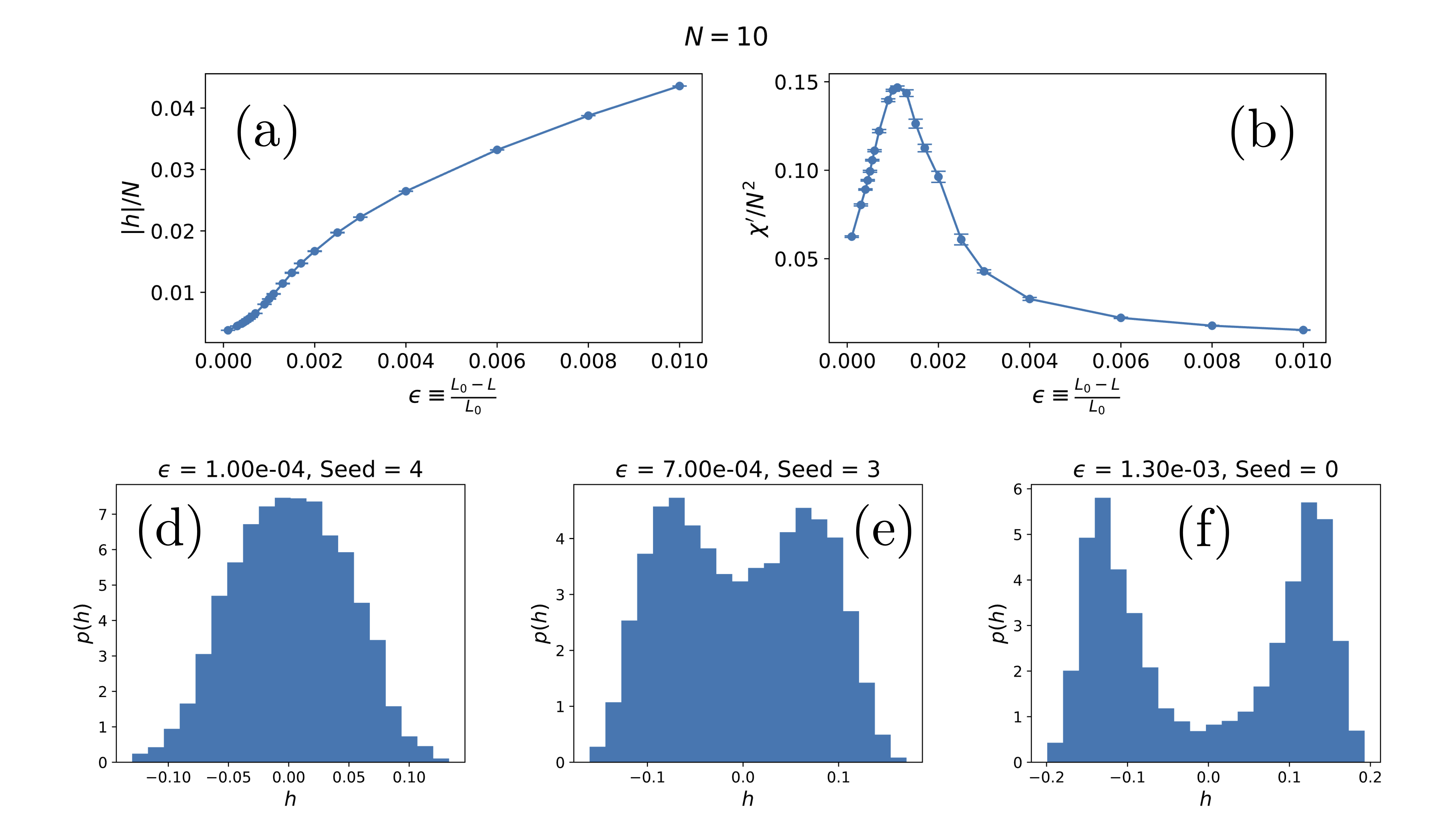}
    \caption{Markov chain Monte Carlo results for $N=10$. Error bars in the line plots represent the standard error of the mean of $10$ independent trials. (a) $\left<|h|\right>/N$ vs $\epsilon$. The order parameter increases as the polymer is compressed into the buckled phase. (b) $\chi'/N^2$ vs $\epsilon$. There is a characteristic peak at the critical value of the compression. (c-e) Probability distribution $p(h)$ of the order parameter, for increasing strains and various seed configurations. As the compression increases from (c) to (e), the distribution transitions from a Gaussian about $h=0$ to a bimodal distribution, indicating a buckling transition.}
    \label{fig:N10_data}
\end{figure*}

\begin{figure*}
    \centering
    \includegraphics[width=0.93\linewidth]{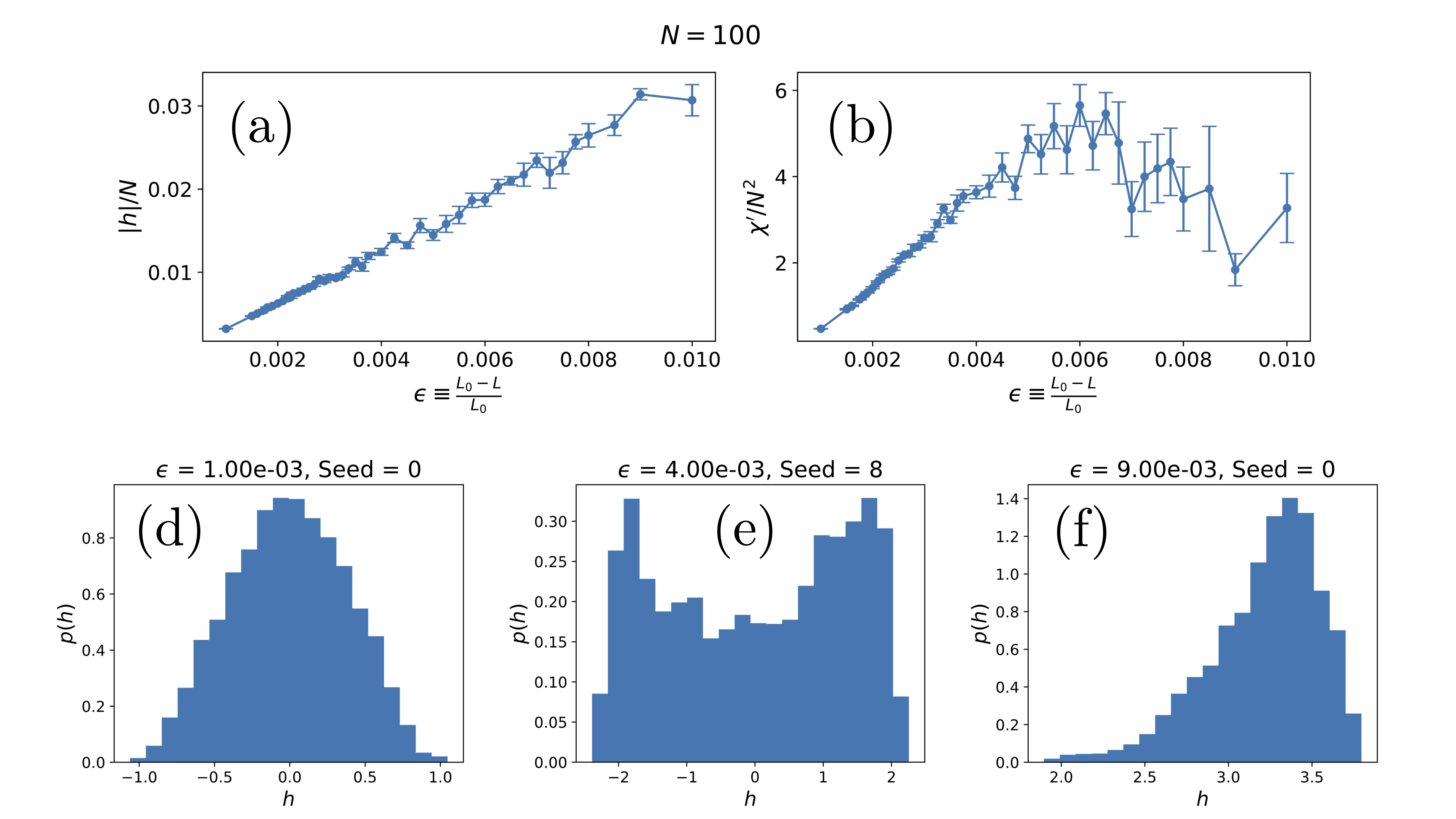}
    \caption{Markov chain Monte Carlo results for $N=100$. Error bars represent the standard error of the mean of $10$ independent trials. (a) $\left<|h|\right>/N$ vs $\epsilon$. (b) $\chi'/N^2$ vs $\epsilon$. Compared to the $N=10$ results of Figure \ref{fig:N10_data}(b), the peak is broader and has shifted to a larger compressive strain. (c-e) Probability distributions $p(h)$ of the order parameter, for increasing strains and various seed configurations.}
    \label{fig:N100_data}
\end{figure*}

Within this scaling range, we vary the compressive strain $\epsilon=(L_0-L)/L_0$ and perform simulations for seven different system sizes, with $N=10, 15, 20, 30, 40, 60$, and $100$. For each system size and imposed strain, we performed $10$ independent trials to generate statistics on measured observables. Runs lasted up to $2\times 10^9$ steps, alternating between local and wave moves, with a snapshot recorded every $2\times 10^5$ steps to avoid correlated samples. The first half of a run was used to allow the system to reach equilibrium, after which data were collected. In Figures \ref{fig:N10_data} and \ref{fig:N100_data}, we compute the average of the absolute value of the order parameter $\left<|h|\right>$ and a susceptibility $\chi'$ (defined below) as a function of the imposed compressive strain. The order parameter is computed by averaging the transverse displacement of the interior nodes:
\begin{equation}
    h=\frac{1}{N-1}\sum_{i=1}^{N-1}(\mathbf{r}_i)_y.
\end{equation}
The susceptibility $\chi$ is related to the variance of the order parameter $h$. For finite system simulations, it is conventional instead to compute the related quantity $\chi'$ given by the variance of the absolute value of the order parameter $|h|$ \cite{binder1997applications}:
\begin{equation}
    \chi'=\frac{N-1}{k_BT}\left(\left<|h|^2\right>-\left<|h|\right>^2\right),
\end{equation}
which has a peak at the buckling transition. Finally, we also compute the probability distributions of the order parameter $p(h)$ at various strains: for small compression, the system is unbuckled and the distribution is Gaussian and centered about $h=0$, but as the compression increases, the spread increases before becoming bimodal in the buckled phase. 

\begin{figure*}
    \centering
    \includegraphics[width=1\linewidth]{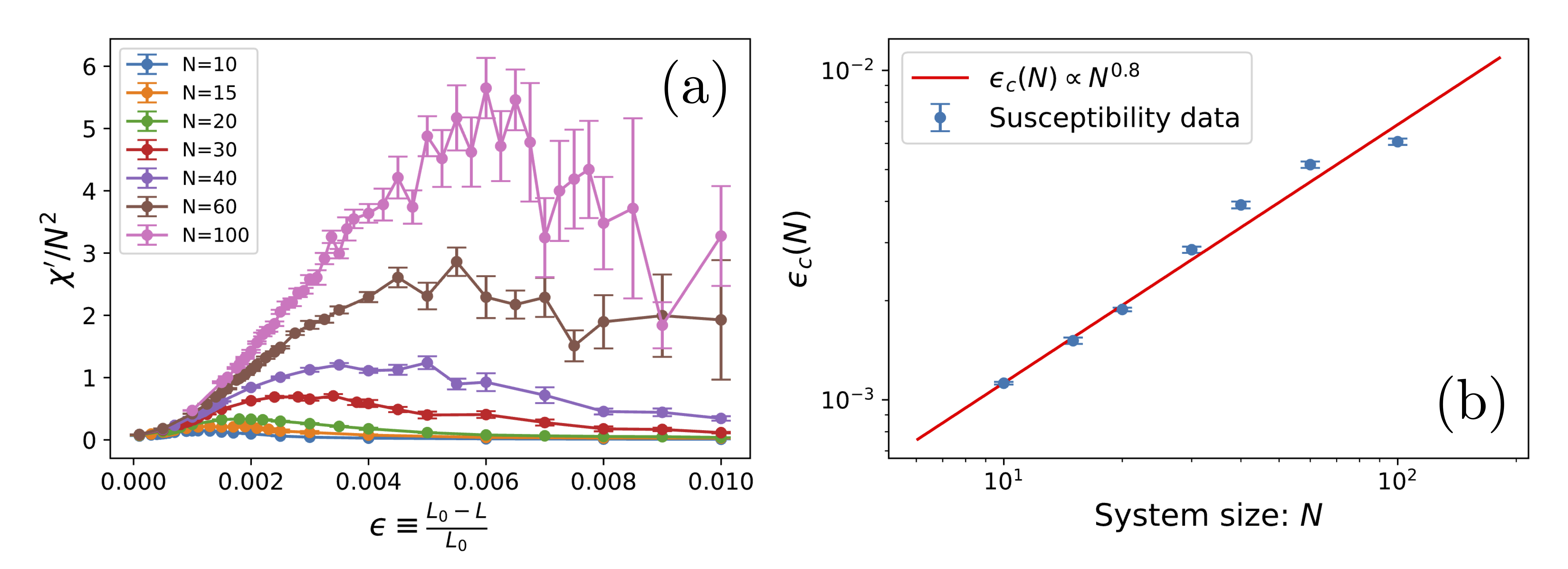}
    \caption{Finite size scaling. (a) The location of the peaks in $\chi'/N^2$ give $\epsilon_c(N)$, which increase with system size. This is in stark contrast to classical buckling, where the critical strain decreases with system size as $N^{-2}$. (b) Critical strain as a function of system size $\epsilon_c(N)$. Estimates are made using the susceptibility data in (a).}
    \label{fig:scaling_data}
\end{figure*}

We can repeat these numerical calculations for the other system sizes. In Figure \ref{fig:scaling_data}a, $\chi'/N^2$ is plotted against the compressive strain for all system sizes. The peak in $\chi'$ indicates the critical strain $\epsilon_c$, and so we see that as the system size $N$ increases, the critical strain $\epsilon_c(N)$ also increases. By fitting the peaks in $\chi'/N^2$ with parabolas, we can estimate $\epsilon_c(N)$, plotted in blue in Figure \ref{fig:scaling_data}b. We find that the critical strain scales roughly as $\epsilon_c(N)\propto N^{0.8}$. The positive exponent is qualitatively consistent with predictions from Section \ref{sec:thermal_fluctuations}, and it highlights how thermal fluctuations lead to a dramatic change in scaling compared to classical buckling, where the critical strain would normally decrease with system size $N$: $\epsilon_c(N) \propto N^{-2}$, see Eq. (\ref{eq:criticalpt}). 

\begin{figure*}
    \centering
    \includegraphics[width=1\linewidth]{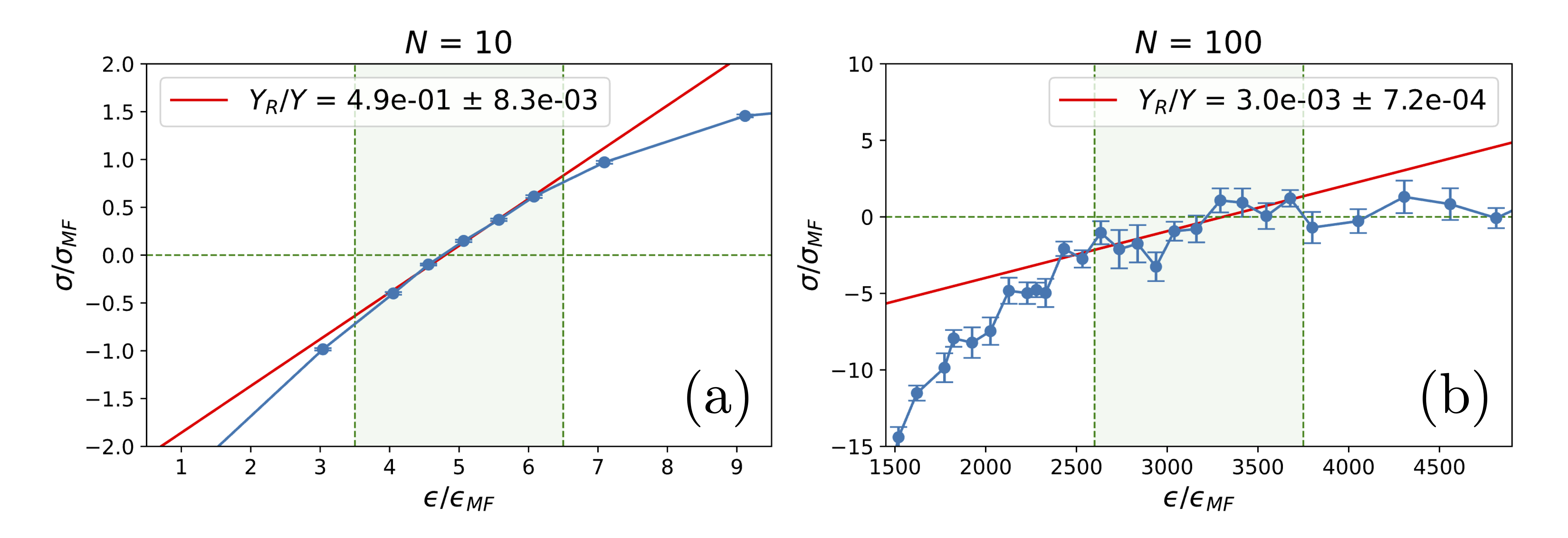}
    \caption{Stress-strain curves for (a) $N=10$ and (b) $N=100$. The $y$-axis is the stress normalized by the mean-field prediction for the critical stress: $\sigma_{\rm MF}=\kappa(\pi/L)^2$. The $x$-axis is the strain normalized by the mean-field prediction for the critical strain: $\epsilon_{\rm MF}=\frac{\kappa}{Y}(\pi/L)^2$. The critical strain is estimated by where the virial stress vanishes, and the renormalized Young's modulus is estimated by the slope of the tangent at that point. Data points within the green region were used to fit the red tangent line.}
    \label{fig:stress_strain}
\end{figure*}

We estimated the critical strain $\epsilon_c(N)$ by locating the peak in the susceptibility, which is a standard perspective from statistical physics. But as a consistency check, it is useful to also estimate the critical strain from stress-strain curves, which can be obtained by computing the virial stress of the system. The virial stress for a system with two-body potentials is given by \cite{frenkel2023understanding}
\begin{equation}
\sigma_{xx}=-\frac{1}{2L}\sum_{k,l}\mathbf{r}_x^{(kl)}\mathbf{f}_x^{(kl)},
\end{equation}
where $\mathbf{r}_x^{(kl)}=\mathbf{r}_x^{(k)}-\mathbf{r}_x^{(l)}$ is the $x$-component of the relative displacement between particles $k, l$ and $\mathbf{f}_x^{(kl)}$ is the $x$-component of the force on particle $k$ due to particle $i$. While this formula can be applied straightforwardly to the stretching energy, the bending energy is a three-body potential, which can be handled \cite{chen2006local,fu2014computing} by noting that the total force $\mathbf{f}_k$ on an individual particle $k$ can be decomposed into a sum of central forces $\mathbf{f}^{(kl)}$, since the energy is simply a function of the separations $E(\{r_{ij}\})$. This procedure gives
\begin{equation}
\begin{split}
&\mathbf{f}_k=-\frac{\partial E}{\partial \mathbf{r}_k}=-\sum_{l\neq k}\frac{\partial E}{\partial r_{kl}}\frac{\partial r_{kl}}{\partial \mathbf{r}_k}=\sum_{l\neq k}\left(-\frac{\partial E}{\partial r_{kl}}\right)\hat{\mathbf{r}}_{kl}\\
&\mathbf{f}^{(kl)}=-\frac{\partial E}{\partial r_{kl}} \hat{\mathbf{r}}_{kl}.
\end{split}
\end{equation}
The resulting stress-strain curves for system sizes $N=10$ and $N=100$ are plotted in Figure \ref{fig:stress_strain}.

\begin{figure*}
    \centering
    \includegraphics[width=1\linewidth]{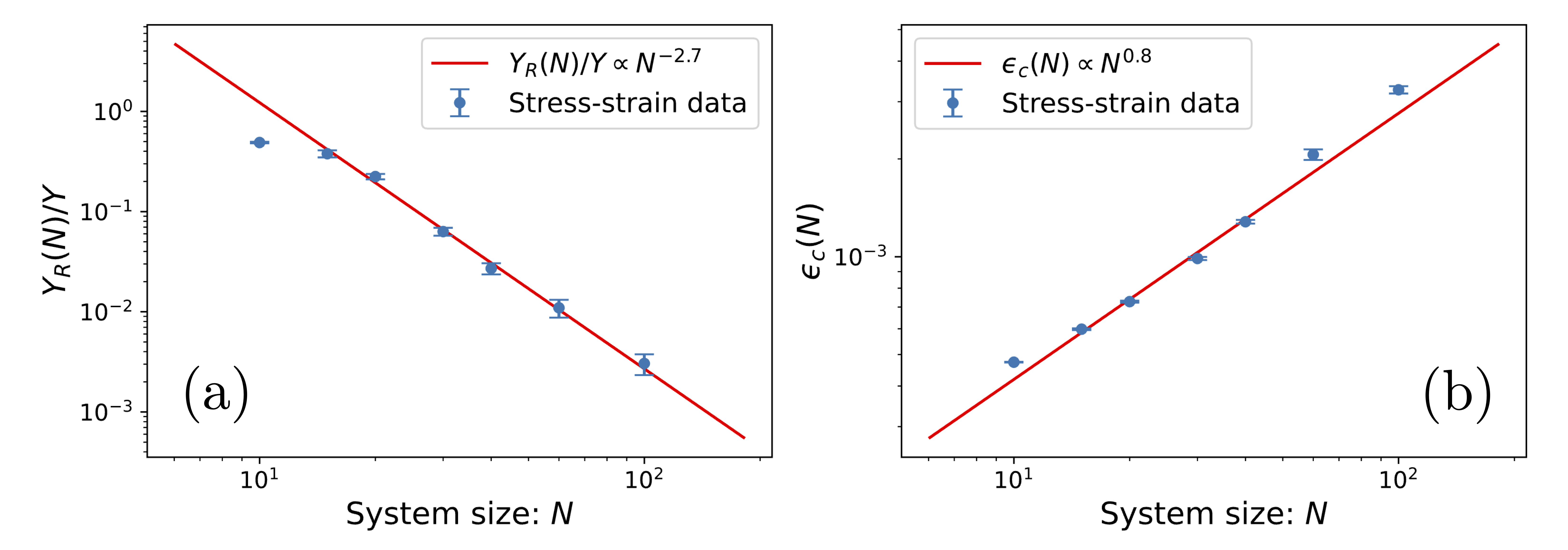}
    \caption{Finite size scaling using data from the stress-strain curves. Fits are across the six largest system sizes. (a) Scale-dependent softening of the renormalized Young's modulus $Y_R(N)/Y$. (b) Critical strain as a function of system size $\epsilon_c(N)$. Results are roughly consistent with the susceptibility-based measurements in Figure \ref{fig:scaling_data}, as well as the heuristic in Eq. (\ref{eq:criticalpt_R}).}
    \label{fig:YR_scaling}
\end{figure*}

In the absence of thermal fluctuations, the stress-strain curve should display a characteristic post-buckling shoulder \cite{hanakata2021thermal}. We see that thermal effects partly smooth out this shoulder, but we can roughly estimate the critical strain by where the virial stress vanishes. Furthermore, estimating the slope of the stress-strain curve at this point gives the renormalized Young's modulus $Y_R$ \cite{hanakata2021thermal}. After repeating these numerical calculations for all seven system sizes, we can measure the size-dependent softening of the renormalized Young's modulus, shown in Figure \ref{fig:YR_scaling}a. The largest six system sizes fall roughly on a power law, with a scaling exponent indicating softening: $Y_R(N)/Y \propto N^{-2.7}$. As shown in Figure \ref{fig:YR_scaling}b, we see that the critical strain for thermal buckling increases with system size: $\epsilon_c(N)\propto N^{0.8}$, consistent with our susceptibility based estimates. Additionally, we note that the scaling of the softening of $Y_R$ and the increase of $\epsilon_c$ are roughly consistent with the following heuristic estimate of replacing the elastic constants in the classical critical strain in Eq. (\ref{eq:criticalpt}) with their renormalized forms:
\begin{equation} \label{eq:criticalpt_R}
    \epsilon_c(L) \sim \frac{\kappa}{Y_R(L)}\left(\frac{\pi}{L}\right)^2,
\end{equation}
where $L\sim N$ and the bending rigidity $\kappa$ does not renormalize in a scale-dependent way. Finally, we point out that obtaining estimates of other scaling exponents is challenging within conventional finite size scaling. Traditional critical phenomena have a well-defined thermodynamic limit, so that the system behaves scale-free beyond the Ginzburg length scale $\ell_{th}$, and in a finite size $L$, the physics is only controlled by the ratio $L/\ell_{th}$. This feature forms the basis of conventional finite size scaling. But mechanical instabilities such as the buckling of 1D polymers are unconventional for multiple reasons. First, the critical point is strongly system size dependent, leading to a breakdown of standard finite size scaling (see \cite{shankar2021} for an example of such a failure in the buckling of 2D sheets). Second, 1D polymer buckling has a second macroscopic length scale, the persistence length $\ell_p$. As a result, the conventional thermodynamic limit $L\to\infty$ cannot be simply taken at fixed persistence length $\ell_p$, and the physics of the polymer beyond the Ginzburg length depends on two length scale ratios: $L/\ell_{th}$ and $L/\ell_p$. A nontraditional scaling limit where $L\to\infty$ while the ratio $L/\ell_p$ is held fixed to a value less than one is possible and likely necessary to observe many of the novel exponents predicted, but this regime is challenging to probe in simulations as it requires increasingly stiff polymers that are difficult to equilibrate. In Appendix \ref{app:binder_cumulant} we describe some of these issues that arise when standard finite size scaling methods are employed for our problem.

\section{Discussion} \label{sec:conclusions}

In this paper, we studied how thermal fluctuations affect the buckling transition of semiflexible linear polymers in an isometric (fixed strain) mechanical ensemble. First, we found that thermal fluctuations shift the critical buckling compression to a higher value compared to zero temperature. Furthermore, the critical strain for thermal buckling \textit{increased} with system size, in contrast to the classical, athermal case. This result also appeared in Monte Carlo simulations of a discretized model of the system. Next, we used a Ginzburg-like criterion in a perturbative calculation to estimate a length scale $\ell_{\rm{th}} \sim (\kappa^2/Y k_BT)^{1/3}$ above which thermal fluctuations lead to significant softening of the Young’s modulus. Finally, we found that polymers of size $\ell_{\rm{th}} < L < \ell_p$ have non-classical scaling exponents near the thermal buckling transition. This was demonstrated using a momentum shell renormalization group, where we found a non-trivial fixed point corresponding to thermal isometric buckling that had critical exponents distinct from classical Euler buckling. Our simulations focused on checking the anomalous scaling of the critical strain and softening of the effective Young’s modulus, and the prediction of these two non-classical scaling phenomena should be observable in a number of systems, such as microtubules, F-actin, and carbon nanotubes (see Table \ref{tab:length_scales}). To obtain the remaining scaling exponents, it is likely necessary to introduce a nontraditional finite size scaling limit where the system size $L$ increases while the ratio $L/\ell_p$ is fixed. Our results have focused on the statics of thermal buckling polymer at equilibrium, but in future work it would be interesting to study the relaxational dynamics \cite{hallatschek2005propagation, hallatschek2007tensionI, hallatschek2007tensionII, obermayer2007stretching, hallatschek2004overdamped} near the thermal buckling transition as well.

\section{Acknowledgments}

We would like to thank Paul Hanakata, Jonathan Bauermann, Alexander Grosberg, Randall Kamien, and Xiaoming Mao for helpful discussions. This work was supported in part by the National Science Foundation, through the Harvard University Materials Research Science and Engineering Center, grant No. DMR-2011754.

\section{Data availability}

The numerical code that produced the data that support the findings of this article is openly available \cite{code}.

\appendix

\onecolumngrid

\section{Alternative derivation of free energy}
\label{app:free_energy}

In this appendix, we show how the form of the free energy in Eq. (\ref{eq:F}), which was represented solely in terms of the transverse polymer deformations, can also be obtained starting from a continuum elasticity model with tranverse and parallel phonon deformations, a discretized version of which was displayed in Eq. (\ref{eq:discrete_energy}).

In the semiflexible regime, the polymer configuration $\mathbf{r}(x)$ can be described using a Monge representation as a deviation from a straight reference state $\mathbf{r}_0 = x\hat{x}$. The internal coordinate $x$ ranges from $0$ to $L_0$, where $L_0$ is the unstretched length of the polymer. The deviation of the polymer configuration from the reference state is represented using a scalar displacement field $u(x)$ describing deformations parallel to the reference state, as well as a field $\mathbf{h}(x)$ describing deformations transverse to the reference state:
\begin{equation} \label{eq:displacements}
    \mathbf{r}(x)=\mathbf{r}_0+\begin{pmatrix} u(x) \\ \mathbf{h}(x) \end{pmatrix}.
\end{equation}
In general, $\mathbf{h}(x)$ is a vector field with $d_c$ components, where the codimension $d_c$ is $1$ or $2$ when the polymer is confined to a plane or 3d space, respectively. From Eq. (\ref{eq:displacements}), the strain $u_{xx}$ is given by $\dd r^2=\dd r_0^2+2u_{xx}\dd x^2$:
\begin{equation}
    u_{xx}=\frac{\dd u}{\dd x}+\frac{1}{2}\left(\frac{\dd u}{\dd x}\right)^2+\frac{1}{2}\left(\frac{\dd\mathbf{h}}{\dd x}\right)^2,
\end{equation}
and we can write down the free energy as
\begin{equation}
    F[u,\mathbf{h}]=\frac{\kappa}{2}\int_0^{L_0}\left(\frac{\dd^2\mathbf{h}}{\dd x^2}\right)^2 \dd x + \frac{Y}{2}\int_0^{L_0}\left[\frac{\dd u}{\dd x}+\frac{1}{2}\left(\frac{\dd u}{\dd x}\right)^2+\frac{1}{2}\left(\frac{\dd\mathbf{h}}{\dd x}\right)^2\right]^2 \dd x.
\end{equation}
Often, one neglects the $(\dd u/\dd x)^2$ term in the strain, keeping only the leading order $\dd u/\dd x$ contribution. In this case, the free energy is Gaussian in $u$, which can be integrated out in the partition function to obtain an effective free energy like Eq. (\ref{eq:F}), up to a missing $\int (\dd\mathbf{h}/\dd x)^4\, \dd x$ term.  

However, if we keep the $(\dd u/\dd x)^2$ term in the strain, we see that a $\int (\dd\mathbf{h}/\dd x)^4\dd x$ term emerges within a mean-field-like approximation. It is useful to introduce a fluctuating part for the $q\neq 0$ Fourier modes of the strain
\begin{equation}
    \delta\epsilon(x)= \frac{\dd u}{\dd x}+\frac{1}{2}\left(\frac{\dd \mathbf{h}}{\dd x}\right)^2.
\end{equation}
The stretching part $F_s$ of the free energy is then
\begin{equation} \label{eq:F_stretch}
\begin{split}
    F_s=\frac{Y}{2}\int_0^{L_0}\left[\delta\epsilon+\frac{1}{2}\left(\delta\epsilon-\frac{1}{2}\left(\frac{\dd\mathbf{h}}{\dd x}\right)^2\right)^2\right]^2 dx &= \frac{YL_0}{2}\left[\epsilon_0+\frac{1}{2}\epsilon_0^2+\frac{1}{2L_0}\int_0^{L_0}\left(\frac{\dd\mathbf{h}}{\dd x}\right)^2\dd x\right]^2 \\
    & \quad+\frac{Y}{2}\int_0^{L_0}\left[\delta\epsilon+\frac{1}{2}\delta\epsilon^2-\frac{1}{2}\delta\epsilon\left(\frac{\dd\mathbf{h}}{\dd x}\right)^2+\frac{1}{8}\left(\frac{\dd \mathbf{h}}{\dd x}\right)^4\right]^2 \dd x, \\
\end{split}
\end{equation}
where the first term comes from the zero mode of the strain ($\epsilon_0$ is the imposed strain) and the second term from the $q\neq0$ modes. At the Gaussian level, the free energy $F_0[\delta \epsilon,\mathbf{h}]$ is
\begin{equation}
    F_0[\delta\epsilon,\mathbf{h}]=\frac{\kappa}{2}\int_0^{L_0}\left(\frac{\dd^2\mathbf{h}}{\dd x^2}\right)^2 \dd x + \frac{Y}{2}\int_0^{L_0}(\delta\epsilon)^2 \dd x,
\end{equation}
so we can make an equipartition estimate
\begin{equation} \label{eq:equipartition}
    \expval{\delta\epsilon^2}_0 \sim \frac{k_BT}{YL_0}, \quad
    \expval{h^2}_0 \sim d_c\int_{q>1/L_0} \frac{k_BT}{\kappa q^4} \sim \frac{d_ck_BTL_0^3}{\kappa}.
\end{equation}
We can now employ a mean-field-like approximation on the full nonlinear stretching, replacing $\delta\epsilon^2$ in Eq. (\ref{eq:F_stretch}) with its average value from Eq. (\ref{eq:equipartition}) to obtain an effective stretching energy in terms of $\mathbf{h}$ only. We indeed find a $\int (\dd\mathbf{h}/\dd x)^4 dx$ term which emerges in the effective energy, similar to Eq. (\ref{eq:F}) of the main text:
\begin{equation}
    F_s = -\frac{Y\expval{\delta\epsilon^2}_0}{2}\int\left(\frac{\dd\mathbf{h}}{\dd x}\right)^2 dx + \frac{Y\expval{\delta\epsilon^2}_0}{8}\int\left(\frac{\dd \mathbf{h}}{\dd x}\right)^4 \dd x + \frac{Y}{8L_0}\left[\int\left(\frac{\dd\mathbf{h}}{\dd x}\right)^2\dd x\right]^2 + \cdots
\end{equation}

\section{Renormalization group (\texorpdfstring{$\varepsilon$}{Epsilon}-expansion)} \label{app:eps}

In this appendix, we give a renormalization group treatment of an abstract generalization of the buckling polymer problem in the main text. Linear polymers are modeled as a curve $\mathbf{h}(x)$ (where the vector $\mathbf{h}$ has $d_c$ components) with one spatial internal degree of freedom. Let us generalize the polymer system such that it has $D=4-\varepsilon$ internal spatial dimensions instead, where we assume $\varepsilon$ to be small, similar to Ref. \cite{aronovitz1988fluctuations}: $\mathbf{h}(x_1,\cdots,x_D)$. Even when we increase the value of $D$ beyond the physical value of $D=1$, we model the in-plane elasticity as being governed solely by a bulk modulus $B$. In this case, we have the following, generalized elastic energy:
	\begin{equation} 
	\begin{split}
	F[\mathbf{h}]=&\frac{\kappa}{2}\int \left(\nabla^2\mathbf{h}\right)^2 \dd^Dx-\frac{f}{2}\int \left(\nabla\mathbf{h}\right)^2\dd^Dx\\
	&+\frac{g}{8}\int \left(\nabla\mathbf{h}\right)^4 \dd^Dx+\frac{B}{8L^D}\left[\int\left(\nabla\mathbf{h}\right)^2\dd^Dx\right]^2.
	\end{split}
	\end{equation}
In the physical dimension $D=1$, $B=y=Y(1-\epsilon)$ and we recover the original elastic energy in Eq. (\ref{eq:F_rg}). While this generalized problem may be unconventional in the context of elasticity theory, this problem can be mapped to the problem of compressible magnets \cite{rudnick1974}. 

The point of introducing this abstract model is that by working with a number of internal dimensions $D=4-\varepsilon$ such that $\varepsilon$ is small, we can systematically control the higher order nonlinearities that would appear in the gradient expansion of the free energy. If $\varepsilon$ is small, then simple power counting tells us that any further higher order terms are irrelevant, which justifies our truncation of the gradient expansion of the free energy. We now carry out a momentum-shell renormalization group to show that a non-trivial fixed point corresponding to thermal isometric buckling found in Sec. \ref{sec:rg} persists in this controlled setting.  

\begin{figure}
	\centering
	   \includegraphics[width=\linewidth]{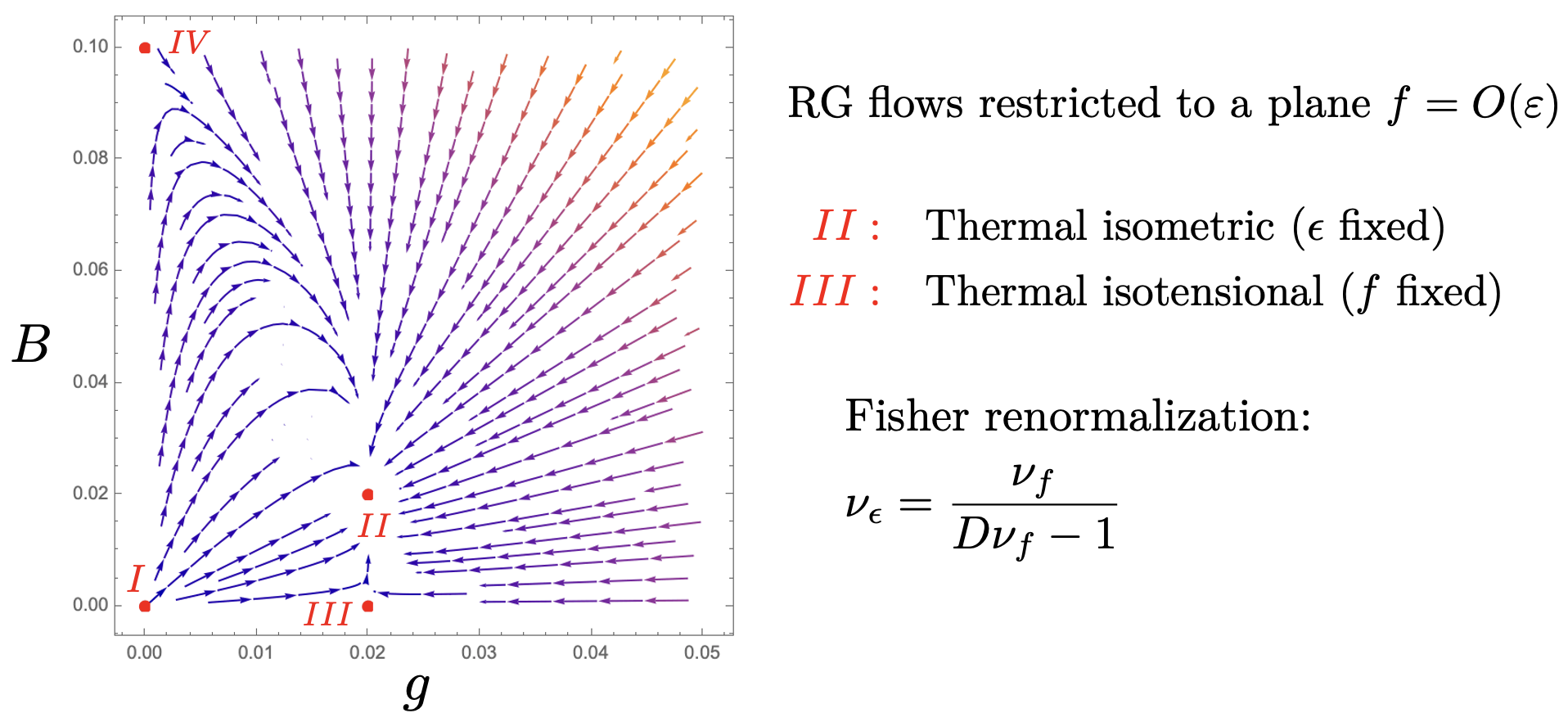} 
	   \caption{RG flows restricted to a plane of $f=O(\epsilon)$, for parameters $\varepsilon=0.1$ and $d_c=2$}
    \label{fig:rg_flows_eps}
	\end{figure}

After using the same standard procedure as outlined in Section \ref{sec:rg} of the main text, we can evaluate corrections to the propagator and two quartic vertices that are lowest order in $\varepsilon$ and obtain the recursion relations
\begin{equation} \label{eq:eps_RG_recursion}
	\begin{split}
	\frac{\dd\kappa}{\dd l}&=(2\zeta+D-4)\kappa,\\
	\frac{\dd f}{\dd l}&=(2\zeta+D-2)f-\frac{d_c B+(d_c+2)g}{2}\cdot \frac{K_D \Lambda^Dk_BT}{\kappa \Lambda^2-f},\\
	\frac{\dd B}{\dd l}&=(4\zeta+D-4)B-B\cdot \frac{d_cB+2(d_c+2)g}{2}\cdot \frac{K_D \Lambda^Dk_BT}{(\kappa \Lambda^2-f)^2},\\
	\frac{\dd g}{\dd l}&=(4\zeta+D-4)g-\frac{(d_c+8)g^2}{2}\cdot \frac{K_D \Lambda^Dk_BT}{(\kappa \Lambda^2-f)^2},
	\end{split}
	\end{equation}
where $K_D=S_D/(2\pi)^D$ and $S_D=2\pi^{D/2}/\Gamma(D/2)$ is the surface area of a unit sphere in $D$ dimensions. For $D=1$, we recover the recursion relations in Eq. (\ref{eq:RG_recursion}) of the main text. In $D=4-\varepsilon$ dimensions, after using a rescaling factor $\zeta=\frac{4-D}{2}$ to keep $\kappa$ fixed in Eq. (\ref{eq:eps_RG_recursion}), we get the recursion relations
\begin{equation}
	\begin{split}
	\frac{\dd f}{\dd l}&=2f-\frac{d_c B+(d_c+2)g}{2}\cdot \frac{1}{1-f},\\
	\frac{\dd B}{\dd l}&=\varepsilon B-B\cdot \frac{d_cB+2(d_c+2)g}{2}\cdot \frac{1}{(1-f)^2},\\
	\frac{\dd g}{\dd l}&=\varepsilon g-\frac{(d_c+8)g^2}{2}\cdot \frac{1}{(1-f)^2},
	\end{split}
	\end{equation}
where we made the simplifying change of variables $f/\kappa \Lambda^2\to f$, $BK_4k_BT/\kappa^2\to B$, and $gK_4k_BT/\kappa^2\to g$. Note that for $d_c=1$, these recursion relations are equivalent to those of the compressible Ising model \cite{rudnick1974}. To leading order in $\varepsilon$, we find four fixed points, as shown in Table \ref{tab:eps_fixed_pts} and plotted in Figure \ref{fig:rg_flows_eps}.
\begin{table}
\caption{\label{tab:eps_fixed_pts}
$O(\varepsilon)$ fixed points}
\begin{ruledtabular}
\def\arraystretch{1.9}
\begin{tabular}{ccc}
Fixed point  & Interpretation & Strain-like \\
$(f^*,B^*,g^*)$ & $(d_c<4)$ & Eigenvalue \\
\hline
    $(0,0,0)$ & Classical Euler & $2$ \\
	\hline
	$\left(\frac{3\varepsilon}{d_c+8},\frac{2(4-d_c)\varepsilon}{d_c(d_c+8)},\frac{2\varepsilon}{d_c+8}\right)$ & Thermal isometric & $2-\frac{6\varepsilon}{d_c+8}$ \\
	\hline
	$\left(\frac{(d_c+2)\varepsilon}{2(d_c+8)},0,\frac{2\varepsilon}{d_c+8}\right)$ & Thermal isotensional & $2-\frac{d_c+2}{d_c+8}\varepsilon$ \\
	\hline
	$(\frac{\varepsilon}{2},\frac{2\varepsilon}{d_c},0)$ & & $2-\varepsilon$ \\
\end{tabular}
\end{ruledtabular}
\end{table}
If we consider physical codimensions ($d_c<4$), fixed point II corresponds to the thermal buckling in the isometric ensemble. It is stable to perturbations in the $g$ and $B$ directions, while the eigenvalue corresponding to the strain is relevant. It is found to be 
\begin{equation}
\lambda=2-\frac{6\varepsilon}{d_c+8}+O(\varepsilon^2),
\end{equation}
so the critical exponent $\nu=1/\lambda$ is
\begin{equation}
\nu=\frac{1}{2}+\frac{3\varepsilon}{2(d_c+8)}+O(\varepsilon^2).
\end{equation}
If we extrapolate to $\varepsilon=3$, this gives $\nu(d_c=1)=1$ and $\nu(d_c=2)=19/20$. 
Similar to the fixed dimension calculation in Section \ref{sec:rg}, there is an invariant plane $g=\frac{d_c}{4-d_c}B$ that is an attractive subspace within the three dimensional parameter space of flows. By computing the irrelevant eigenvalue at the thermal isometric fixed point, we can also estimate the softening exponent of the Young's modulus: $Y_R/Y\sim (L/\ell_{\rm{th}})^{-\varepsilon}$. This means $\eta_u=\varepsilon$, and since $\eta=0$, the Ward identity $2\eta+\eta_u=4-D$ holds to first order in $\varepsilon$, and so the scaling relations for thermally buckling polymers should be the same as those derived for sheet polymers in \cite{shankar2021}, at least to first order in $\varepsilon$:
\begin{equation}
    \beta = \nu\left(1-\frac{\eta}{2}\right), \quad \gamma=2\beta, \quad \delta=3.
\end{equation}

The critical exponent values to first order in $\varepsilon$ are not expected to be numerically accurate for the physically relevant value of $\varepsilon=3$, but the qualitative behavior is expected to hold, and the important point is that there is a new, interacting fixed point that governs the scaling behavior near the buckling transition for polymers of size $\ell_{\rm{th}}<L<\ell_p$. The $\varepsilon$-expansion thus gives a consistency check for the results of Section \ref{sec:rg} in a more controlled context. 

Fixed point III, which satisfies $B^*=0$, has a physical interpretation of a thermal buckling polymer in an \textit{isotensional} ensemble where a fixed force $f$ is applied at the ends, which are otherwise free to move. An example of isotensional buckling is in the experiments where microtubules were buckled inside vesicles \cite{elbaum1996}. The buckling transition is well-defined in the case of a semiflexible polymer, as the rod-like elasticity ensures there is a flat phase and a buckled phase. Mathematically, this is captured within the fourth order nonlinear $g$ term in the free energy. In the isotensional ensemble, the nonlocal $B$ interaction is identically zero, as this term only emerges due to the constraint of fixed endpoints. Within the renormalization group flows, $B=0$ is an invariant plane, and this subspace captures the physics of isotensional thermal buckling, with yet another set of critical exponents for fixed point III that can be worked out from the RG recursion relations described above. 

This situation is reminiscent of thermal buckling in isotropically compressed \textit{sheets} \cite{shankar2021}. Since the isometric and isotensional ensembles are thermodynamically conjugate ensembles, one might assume that the physical scaling properties of the two ensembles are the same in the thermodynamic limit. However, thermal buckling is an example of Fisher renormalization \cite{fisher1968}, where the critical exponents in the two ensembles are distinct, which is possible due to the diverging correlation length at a critical point. Previous work inspired by single-molecule biophysics has appreciated the inequivalence of mechanical ensembles due to finite-size effects \cite{lubensky2002single, keller2003relating, sinha2005inequivalence, dutta2018inequivalence, dutta2020statistical}, but the inequivalence of ensembles discussed here would persist even in an idealized thermodynamic limit. 

Moreover, there are scaling relations between the critical exponents in the isometric (denoted by a subscript $\epsilon$) and isotensional (denoted by a subscript $f$) ensembles \cite{shankar2021}. To order $O(\varepsilon)$ within the $\varepsilon=4-D$ expansion, we verify that the Fisher relations between the critical exponents in the two ensembles hold the case of buckling linear polymers:
\begin{equation}
\nu_\epsilon=\frac{\nu_f}{D\nu_f-1}.
\end{equation}

Despite the similarities outlined above, there are also differences between the thermal buckling of sheets versus polymers. In the case of sheets, all exponents in both ensembles were determined by a single exponent $\eta$ \cite{shankar2021}. However, this is not the case for semiflexible polymers. In the case of isotensional sheets, there is no renormalization of the tension, and this observation led to additional scaling relations. On the other hand, for isotensional linear polymers, the origin of the nonlinear term is in higher order stretching terms that couple to the symmetry-breaking applied force, and so there is no analogous simplification. 

\section{Saddle point estimate of the critical strain}

\label{app:saddlept}

In this appendix, we will work with the full nonlinear elastic free energy from Eq. (\ref{eq:F}) and use a saddle point approximation to estimate the thermal-induced shift in the critical buckling strain $\epsilon_c$. It is helpful to rewrite the free energy in Eq. (\ref{eq:F}): recall that $L_0=\frac{L}{1-\epsilon}$, and define both $y=Y(1-\epsilon)$ and $A_0=\frac{1}{L}\int_0^L\frac{1}{2}\left(\frac{\dd\mathbf{h}}{\dd x}\right)^2\dd x$. We will use the bar notation ($\overline{\kappa}=\kappa/k_BT, \overline{f}=f/k_BT, \overline{y}=y/k_BT$) to designate a factor of $1/k_BT$ that occurs in the Boltzmann distribution. The free energy in Eq. (\ref{eq:F}) is then
\begin{equation} \label{eq:betaF}
\begin{split}
\frac{F[\mathbf{h}]}{k_BT}=\frac{\overline{\kappa}}{2}\int_0^L &\left(\frac{\dd^2\mathbf{h}}{\dd x^2}\right)^2 \dd x-\overline{f}LA_0\\
&+\frac{\overline{f}}{8}\int_0^L\left(\frac{\dd\mathbf{h}}{\dd x}\right)^4\dd x+\frac{\overline{y}LA_0^2}{2}.
\end{split}
\end{equation}
The nonlocal quartic nonlinearity is embodied in the factor $A_0^2$ in the final term. We rewrite the partition function $\mathcal{Z}=\int \mathcal{D}\mathbf{h}(x)\, e^{-F[\mathbf{h}]/k_BT}$ into a more useful form by employing a Hubbard-Stratonovich transformation on each of the two quartic terms in Eq. (\ref{eq:betaF}). This introduces two auxiliary variables, a scalar quantity $\sigma$ and a field $\mu(x)$:
\begin{equation}
e^{-\frac{\overline{y}LA_0^2}{2}}=\sqrt{\frac{\overline{y}L}{2\pi}}\int \dd\sigma \; e^{-\frac{\overline{y}L}{2}\sigma^2+i\overline{y}LA_0\sigma},
\end{equation}
\begin{equation}
\begin{split}
e^{-\frac{\overline{f}}{8}\int_0^L\left(\frac{\dd\mathbf{h}}{\dd x}\right)^4\dd x}=\mathcal{N}\int\mathcal{D}\mu(x) e^{-\frac{\overline{f}}{2}\int_0^L \left[\mu(x)^2-i\mu(x)\left(\frac{\dd\mathbf{h}}{\dd x}\right)^2\right]\dd x},
\end{split}
\end{equation}
where $\mathcal{N}$ is a normalization factor. The partition function $\mathcal{Z}$ is now of the form
\begin{equation}
\mathcal{Z}=\mathcal{N}\sqrt{\frac{\overline{y}L}{2\pi}}\int\mathcal{D}\mu(x) \int \dd\sigma \; e^{-\frac{\overline{y}L}{2}\sigma^2-\frac{\overline{f}}{2}\int_0^L \mu(x)^2\dd x}\cdot \mathcal{Z}_{\sigma,\mu},
\end{equation}
where $\mathcal{Z}_{\sigma,\mu}$ is a quadratic function of the transverse displacement field $\mathbf{h}(x)$:
\begin{equation}
\begin{split}
\mathcal{Z}_{\sigma,\mu}&=\int \mathcal{D}\mathbf{h}(x) e^{-F_{\sigma,\mu}[\mathbf{h}]/k_BT},\\
F_{\sigma,\mu}[\mathbf{h}]&=\frac{\kappa}{2}\int_0^L \left(\frac{\dd^2\mathbf{h}}{\dd x^2}\right)^2 \dd x-fLA_0-iyLA_0\sigma \\ 
&\qquad-\frac{if}{2}\int_0^L\mu(x)\left(\frac{\dd\mathbf{h}}{\dd x}\right)^2\dd x.
\end{split}
\end{equation}

To proceed further, we use an ansatz $\mu(x)=\mu$ for the auxiliary field $\mu(x)$, which restricts the full functional integral over the auxiliary field $\mu(x)$ to just its constant mode. We justify this simplification because the effective free energy in the partition function does not involve gradients of $\mu(x)$, so we consider a uniform value for simplicity. After using this ansatz, the partition function takes the simpler form
\begin{equation}
\mathcal{Z}=\mathcal{N}\sqrt{\frac{\overline{y}L}{2\pi}}\int \dd\mu \int \dd\sigma \; e^{-\frac{\overline{y}L}{2}\sigma^2-\frac{\overline{f}L}{2}\mu^2}\cdot \mathcal{Z}_{\sigma,\mu},
\end{equation}
and $\mathcal{Z}_{\sigma,\mu}$ now only involves Gaussian integrations, where the polymer is subject to an effective force of $f+i(\sigma y+\mu f)$:
\begin{equation}
\begin{split}
\mathcal{Z}_{\sigma,\mu}&=\int \mathcal{D}\mathbf{h}(x) \exp\bigg[-\frac{\overline{\kappa}}{2}\int_0^L \left(\frac{\dd^2\mathbf{h}}{\dd x^2}\right)^2 \dd x\\
&\hspace{2cm}+\frac{\overline{f}+i(\sigma\overline{y}+\mu\overline{f})}{2}\int_0^L\left(\frac{\dd\mathbf{h}}{\dd x}\right)^2\dd x\bigg]\\
&=\mathcal{N}'\prod_q\left(\frac{2\pi L}{\overline{\kappa}q^4-(\overline{f}+i\sigma\overline{y}+i\mu\overline{f})q^2}\right)^{d_c}.
	\end{split}
\end{equation}
We now rewrite the partition function using Fourier modes as
\begin{equation}
\mathcal{Z}\propto \int \dd\mu \int \dd\sigma \, e^{-L\mathcal{F}(\sigma,\mu)},
\end{equation}
\begin{equation}
\begin{split}
\mathcal{F}(\sigma,\mu)=&\frac{y}{2k_BT}\sigma^2+\frac{f}{2k_BT}\mu^2\\
&+\frac{d_c}{L}\sum_q \ln[\kappa q^4-(f+i\sigma y+i\mu f)q^2].
\end{split}
\end{equation}
The partition function can be estimated using a saddle-point approximation, where the integrand is dominated by the value of $(\sigma,\mu)$ which minimizes $\mathcal{F}(\sigma,\mu)$:
\begin{equation} \label{eq:saddlept}
\begin{split}
\frac{\partial \mathcal{F}}{\partial \sigma}&=\frac{y\sigma}{k_BT}+\frac{d_c}{L}\sum_q\frac{-iy}{\kappa q^2-(f+i\sigma y+i\mu f)}=0,\\
\frac{\partial \mathcal{F}}{\partial \mu}&=\frac{f\mu}{k_BT}+\frac{d_c}{L}\sum_q\frac{-if}{\kappa q^2-(f+i\sigma y+i\mu f)}=0.
\end{split}
\end{equation}
The saddle point value for $i(\sigma y+\mu f)\equiv-\lambda Y$ represents the fluctuation-induced shift in the critical strain: $\epsilon_c(T)=\epsilon_c(T=0)+\lambda$. Intuitively, $\lambda>0$, as thermal fluctuations lead to transverse deformations that decrease the distance between endpoints when there are free boundary conditions; however, in the isometric ensemble, the endpoints are fixed in place, and this generates a spontaneous tension. Thus, an increased compressive strain is needed to buckle the polymer, as one needs to overcome this tension. By adding the two saddle point constraints in Eq. (\ref{eq:saddlept}), we obtain a simple closed equation for $\lambda$:
\begin{equation}
\frac{L}{k_BT}\cdot \lambda Y-d_c(y+f)\sum_q\frac{1}{\kappa q^2-(f-\lambda Y)}=0.
\end{equation}
It is useful to introduce two dimensionless parameters, which are the persistence parameter $\psi\equiv Lk_BT/\kappa=L/\ell_p$ and a one-dimensional analog of the F\"oppl-von K\'arm\'an number $\text{FvK}\equiv YL^2/\kappa$. Additionally, to evaluate the sum, we consider buckling under pinned boundary conditions, in which case the Fourier basis functions are $\{\sin(n\pi/L):n=1,2,\cdots\}$. This gives 
\begin{equation}
\frac{\lambda}{\psi}-d_c\sum_{n=1}^\infty \frac{1}{(n\pi)^2-\text{FvK}(\epsilon-\lambda)}=0.
\end{equation}
The thermal shift $\lambda$ at the critical point $\epsilon_c=\pi^2/\text{FvK}$ is given by
\begin{equation} \label{eq:shift}
\frac{\lambda}{\psi}=\frac{d_c}{2}\left(\frac{1}{\pi^2-\text{FvK}\cdot\lambda}-\frac{\cot \sqrt{\pi^2-\text{FvK}\cdot\lambda}}{\sqrt{\pi^2-\text{FvK}\cdot\lambda}}\right).
\end{equation}

	\begin{figure*}
	\centering
	   \includegraphics[width=\linewidth]{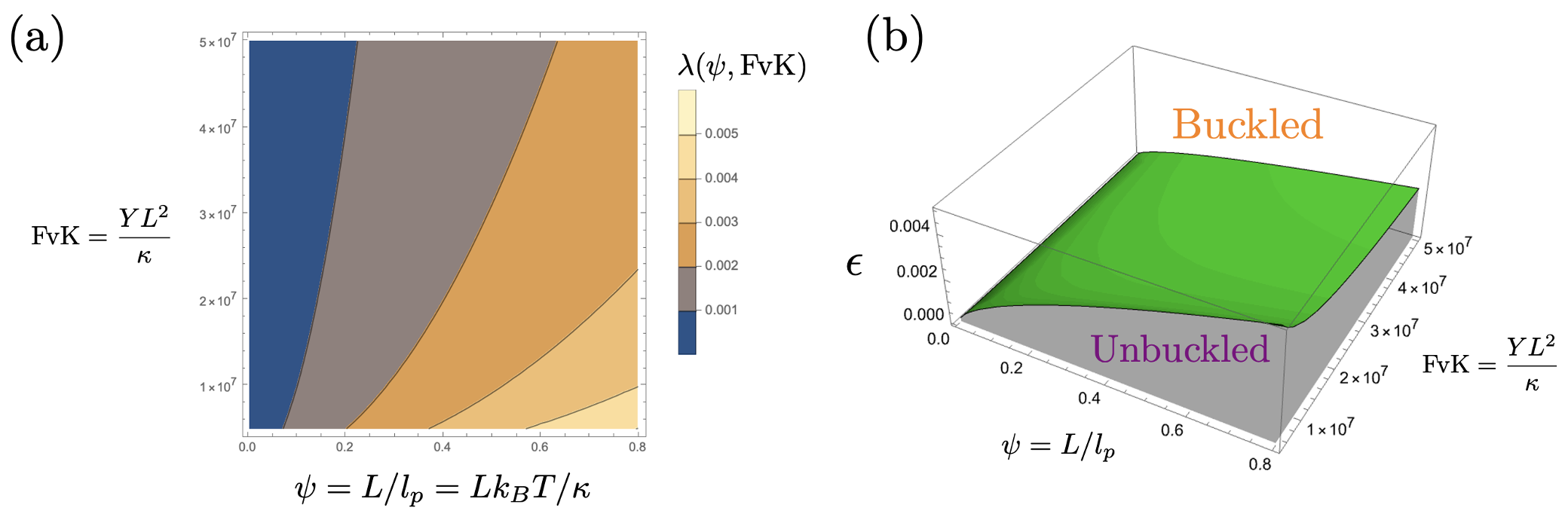} 
	   \caption{Thermal shift $\lambda$ in the critical compression for codimension $d_c=2$ and pinned boundary conditions, calculated within a saddle-point approximation. (a) A contour plot of the thermal shift $\lambda$ as a function of the persistence parameter $\psi=L/\ell_p$ and a 1d analog of the F\"oppl-von K\'arm\'an number $\text{FvK}=YL^2/\kappa$. For context, a microtubule with $L=20\,\mu$m has $\lambda\sim 10^{-4}$. (b) Thermal buckling phase diagram, with a green phase boundary given by $\epsilon_c(\psi,\text{FvK})=\pi^2/\text{FvK}+\lambda(\psi,\text{FvK})$.}
       \label{fig:saddle_shift}
	\end{figure*}

    \begin{figure}
	   \centering
	   \includegraphics[width=0.5\linewidth]{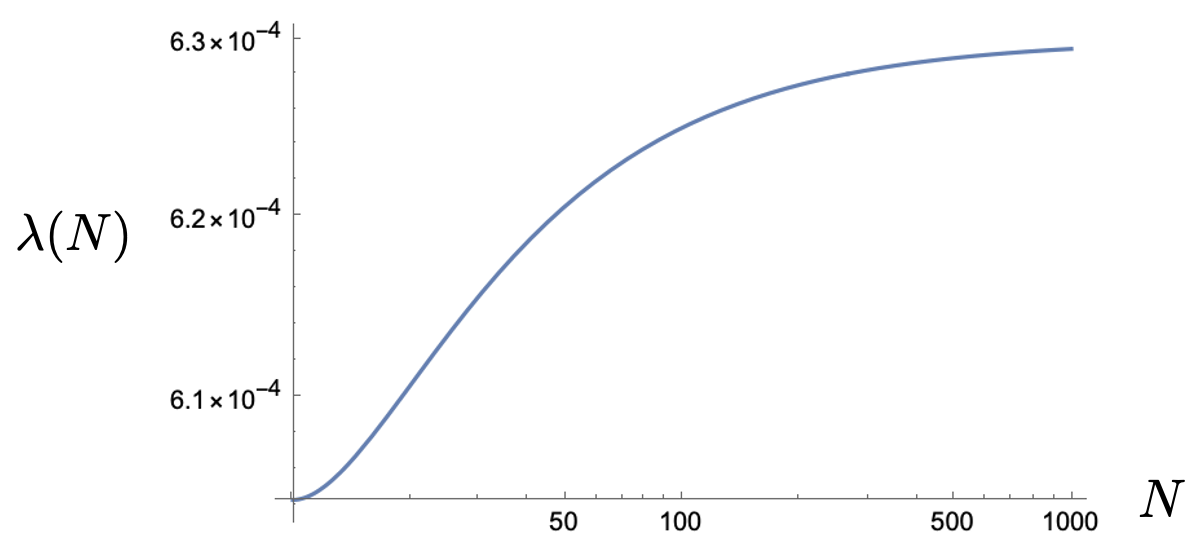} 
	   \caption{Thermal shift from Eq. (\ref{eq:shift}) as a function of the system size $N$ for the simulation parameters in Section \ref{sec:mc} (the thermal length in units of monomer spacings is $N_{th}=(6\pi^4)^{1/3}\approx 10$, and the persistence length is $N_p=1000$).}
        \label{fig:saddle_shift_sim}
	\end{figure}
    
The results are plotted for a variety of $(\psi,\text{FvK})$ in Figure \ref{fig:saddle_shift}. It is particularly relevant to consider the situation where the thermal length and persistence length are fixed, and only the system size is allowed to vary. If we choose elastic constants and a temperature matching those of the Monte Carlo simulations in Section \ref{sec:mc}, we see that the thermal shift $\lambda$ predicted from the saddle point calculation increases with system size, as depicted in Figure \ref{fig:saddle_shift_sim}.

\section{Binder cumulant and finite size scaling}

\label{app:binder_cumulant}

Estimating the scaling exponents of other quantities in Table \ref{tab:exponents} using standard finite size scaling poses challenges. Conventional approaches to finite size scaling assume a thermodynamic limit, and estimate the critical point in the large system size limit $L\to\infty$. This is often done by analyzing the Binder cumulant 
\begin{equation}
    U_L = 1 - \frac{\expval{h^4}}{3\expval{h^2}^2}
\end{equation}
for various system sizes $L$. In conventional finite size scaling, with a critical point $T_c(L)$ that approaches a finite limit as $L\to\infty$, the Binder cumulants of different system sizes will have a common intersection that marks the critical point in the thermodynamic limit \cite{binder1992, binder1997applications}.
    \begin{figure}
	   \centering
	   \includegraphics[width=0.8\linewidth]{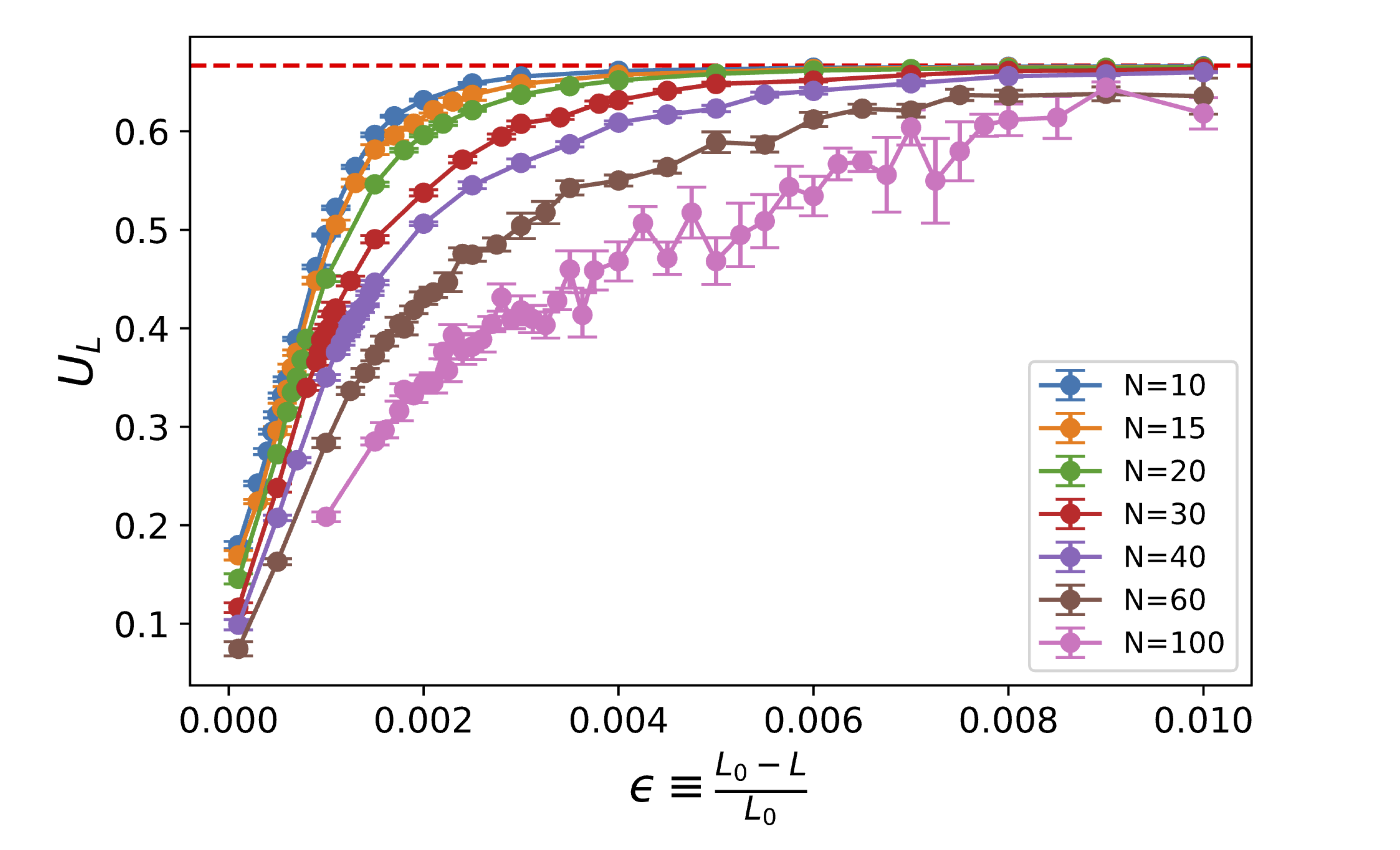} 
	   \caption{Binder cumulants $U_L$ for various system sizes. There appears to be no common intersection, except off at infinity. This would be consistent with the critical strain increasing with system size, as well as the semiflexible polymer description breaking down for sizes beyond the persistence length.}
        \label{fig:binder_cumulants}
	\end{figure}
In Figure \ref{fig:binder_cumulants}, we have plotted the Binder cumulants for various system sizes from the Markov chain Monte Carlo simulations in Section \ref{sec:mc}. We see that there is no common intersection point at finite strain, with the cumulants approaching each other off at infinity. These results are consistent with the results of Section \ref{sec:mc}, where we saw that the critical strain is size dependent, going to infinity for large systems.

\twocolumngrid

\bibliography{refs}

\end{document}